\documentclass[a4paper]{article}
\usepackage[utf8]{inputenc}
\usepackage{amsmath}
\usepackage{graphicx}
\usepackage[colorinlistoftodos]{todonotes}
\usepackage[colorlinks=true, allcolors=blue]{hyperref}
\usepackage[a4paper,top=3cm,bottom=2cm,left=3cm,right=3cm,marginparwidth=1.75cm]{geometry}
\usepackage{amsthm}
\usepackage[ruled,vlined]{algorithm2e}
\usepackage{authblk}
\usepackage{natbib}
\usepackage{setspace}
\theoremstyle{definition}
\newtheorem{definition}{Definition}

\newcommand*{\vertbar}{\rule[-1ex]{0.5pt}{2.5ex}}

\DeclareMathOperator*{\argmin}{\arg\!\min}

\renewcommand{\vec}{\text{vec}}

\newcommand{\beginsupplement}{%
        \setcounter{table}{0}
        \renewcommand{\thetable}{S\arabic{table}}%
        \setcounter{figure}{0}
        \renewcommand{\thefigure}{S\arabic{figure}}%
     }

\allowdisplaybreaks 

\title{Bayesian Multinomial Logistic Normal Models through Marginally Latent Matrix-T Processes}
\date{}
\begin{document}

\author[1,2]{Justin D. Silverman}
\author[1]{Kimberly Roche}
\author[5]{Zachary C. Holmes}
\author[1,4,5,6]{Lawrence A. David}
\author[1,3,4,6]{Sayan Mukherjee}

\renewcommand\Affilfont{\itshape\small}

\affil[1]{Program in Computational Biology and Bioinformatics, Duke University, Durham, NC 27708}
\affil[2]{Medical Scientist Training Program, Duke University, Durham, NC 27708}
\affil[3]{Departments of Statistical Science, Mathematics, Computer Science, Biostatistics \& Bioinformatics, Duke University, Durham, NC 27708}
\affil[4]{Center for Genomic and Computational Biology, Duke University, Durham, NC 27708}
\affil[5]{Department of Molecular Genetics and Microbiology, Duke University, Durham, NC 27708}
\affil[6]{Denotes Co-corresponding Authors}

\maketitle

\newpage
\begin{abstract}
   Bayesian multinomial logistic-normal (MLN) models are popular for the analysis of sequence count data (\textit{e.g.}, microbiome or gene expression data) due to their ability to model multivariate count data with complex covariance structure. However, existing implementations of MLN models are limited to handling small data sets due to the non-conjugacy of the multinomial and logistic-normal distributions. We introduce MLN models which can be written as marginally latent matrix-t process (LTP) models. Marginally LTP models describe a flexible class of generalized linear regression, non-linear regression, and time series models. We develop inference schemes for Marginally LTP models and, through application to MLN models, demonstrate that our inference schemes are both highly accurate and often 4-5 orders of magnitude faster than MCMC.
\end{abstract}

\newpage
\section{Introduction}

In this article we develop methods for efficient inference of Bayesian multinomial-logistic normal (MLN) models. MLN models are used for the analysis of compositions measured through multivariate counting. In contrast to multinomial Dirichlet models, multinomial logistic-normal models permit both positive and negative covariation between multinomial categories \citep{aitchison1980}. While multinomial logistic-normal topic models have been used in natural language processing for some time \citep{Blei2006, Glynn2019}, more recently these models have been adopted for regression and time-series modeling of microbiome data \citep{grantham2017, Silverman2018, aijo2016}. 

Yet, inference in MLN models is difficult due to lack of conjugacy between the multinomial and the logistic normal. Early work with MLN models used Metropolis within Gibbs samplers \citep{cargnoni1997, Billheimer2001} and could scale to just a small number multinomial categories (\textit{i.e.}, less than 5). Recently, P{\'o}lya--Gamma data augmentation was proposed as a means of inference in MLN regression by augmenting P{\'o}lya--Gamma random variables between the multinomial and logistic normal components of a model and sampling each variable as a separate Gibbs sampling step \citep{Polson2013}. Numerous authors have found this approach too computationally intensive to scale to large multinomial models and have instead developed augmentation methods based on a stick-breaking representation of the multinomial  \citep{Linderman2015, Zhang2017}. However, this stick breaking representation does not maintain the logistic-normal form of the model and is sensitive to the labeling of multinomial categories \citep{Linderman2015}. Most recently, 
several authors \citep{Silverman2018,aijo2016,grantham2017} have shown that Hamiltonian Monte Carlo (HMC) provides for a more efficient and scalable approach to inference in MLN models. In particular, \citet{grantham2017} used a HMC within a Gibbs sampler whereas both \citet{Silverman2018}, and \citet{aijo2016} found that the No-U-Turn-Sampling algorithm provided by the Stan Modeling language \citep{Gelman2015}, provided more scalable inference. However, both these approaches are still limited in the number of categories or samples that they can handle. \citet{Silverman2018} analyzed approximately 800 samples each with only 10 multinomial categories; \citet{aijo2016} analyzed 36 multinomial categories but had to run their model over the dataset using a sliding window of 60 samples at a time; and \citet{grantham2017} analyzed 166 samples and 2662 categories but had to impose low rank structure on the logistic normal model for computational tractability. In this work we show that our inference methods scale to hundreds to thousands of categories and samples and permit inference for a wide variety of models including non-linear regression models (as in \citet{aijo2016}), dynamic linear models (as in \citet{Silverman2018}), and linear regression models (as in \citet{grantham2017}).

The key ideas developed in this article are three-fold. First, we demonstrate that a number of popular multivariate Bayesian models are special cases of what we term marginally latent matrix-t process (Marginally LTP) models. As the name suggests, Marginally LTP models share a canonical marginal form which is a generalization of Student-t processes \citep{Jylanki2011, Shah2014} to multivariate problems and alternative observation distributions. Second, through applications to MLN linear models, we demonstrate that the LTP representation can provide efficient inference schemes for Marginally LTP models. In particular our inference is based on collapsing a target model to its LTP form, obtaining samples from that LTP, and then uncollapsing the samples (\textit{i.e.} sampling form a particular posterior conditional) so as to produce samples from the original target model. Finally, we show that for multinomial-logistic normal models, a Laplace approximation to the LTP form can provide efficient and accurate inference for the original target model.

The layout of this article is as follows. In Section 2 we introduce the MLN linear models which motivate subsequent sections. In Section 3 we introduce the class of Marginally LTP models which encompasses MLN linear models. In Section 4 we develop inference methods for Marginally LTP Models. In Section 5 we demonstrate our approaches through an extensive simulation studies of MLN linear models. In Section 6 we demonstrate the utility of MLN linear Models through application to a microbiome sequence count study. Finally, we close with a discussion in Section 7. 
\section{Multinomial Logistic-Normal (MLN) Linear Models} \label{section:multinomial}

Our primary motivation in this work was to develop efficient inference for a class of models we refer to as multinomial logistic-normal (MLN) linear models. Consider a dataset consisting of paired observations $Y$ and $X$ where $Y_{\cdot j}$ denotes the $j$-th $D$-variate count vector with a paired $Q$ dimensional covariate vector $X_{\cdot j}$. For example, in the analysis of microbiome data we may consider $Y_{\cdot j}$ to be the number of $D$ different bacterial taxa sequenced in the $j$-th sample and $X_{\cdot j}$ to be $Q$ different covariates describing the $j$-th sample. We denote the number of samples as $N$. In this setting we define the class of multinomial logistic normal linear models as 
\begin{align}
    Y_{\cdot j} &\sim \text{Multinomial}(\pi_{\cdot j})  \label{eq:mnllinear1}\\
    \pi_{\cdot j} &= \text{ALR}_D^{-1}(\eta_{\cdot j}) \label{eq:mnllinear2}\\
    \eta_{\cdot j} &\sim N(\Lambda X_{\cdot j}, \Sigma) \label{eq:mnllinear3} \\
    \Lambda &\sim N(\Theta, \Sigma, \Gamma)\label{eq:mnllinear4} \\
    \Sigma &\sim IW(\Xi, \upsilon). \label{eq:mnllinear5}
\end{align}
where $ALR^{-1}_D$ is the inverse additive log-ratio transform  given by \citet{aitchison1986}
\begin{equation}
    \text{ALR}_D^{-1}(\eta_{\cdot j})=\left(\frac{e^{\eta_{1j}}}{1+\sum^{D-1}_{i=1}e^{\eta_{ij}}},\cdots,\frac{e^{\eta_{(d-1)j}}}{1+\sum^{D-1}_{i=1}e^{\eta_{ij}}}, \frac{1}{1+\sum^{D-1}_{i=1}e^{\eta_{ij}}}\right). \label{eq:alrdef}
\end{equation}
$ALR^{-1}_D$ is also known as the multivariate-logit and is identical to the softmax transform where one component is set to zero for identifiability. In this way Equations \eqref{eq:mnllinear1} and \eqref{eq:mnllinear2} is equivalent to the multivariate-logit parameterized multinomial commonly used in multinomial regression models. Importantly, the pull-back of the multivariate normal measure through the inverse ALR transform is the logistic-normal measure (\citet{aitchison1980}) and we therefore refer to this class of models as multinomial \textit{logistic-normal} linear models. The definition of $ALR^{-1}$ implies that $\eta$ is a real valued matrix of dimension $(D-1) \times N$, $\Sigma$ is a covariance matrix of dimension $(D-1) \times (D-1)$ and $\Gamma$ is a covariance matrix of dimension $Q \times Q$. Additionally we note that the term $\Lambda \sim N(\Theta, \Sigma, \Gamma)$ denotes a matrix-normal distribution defined by $\vec(\Lambda) \sim N(\vec(\Theta), \Gamma \otimes \Sigma)$ where $\otimes$ denotes the Kronecker product and $\vec$ denotes the vectorization operation that converts a matrix into a column vector.

MLN linear models provide a means of inferring log-linear interactions between compositions measured through multivariate counting and covariates. Each element $\Lambda_{ik}$ describes the linear effect of the $k$-th covariate on the $i$-th log-ratio coordinate. In contrast to standard multinomial logistic regression, the above model assumes that each observation is subject to extra-multinomial variation\footnote{This is reflected in the fact that Equation \eqref{eq:mnllinear3} reads as $\eta_{\cdot j} \sim N(\Lambda X_{\cdot j}, \Sigma)$ rather than $\eta_{\cdot j} = \Lambda X_{\cdot j}$. } and uncertainty regarding the covariation between the rows of $\eta$ and $\Lambda$ (Equation \eqref{eq:mnllinear5}).  Additinoally, the prior specified in Equations \eqref{eq:mnllinear4} and \eqref{eq:mnllinear5} produce shrinkage of estimates of $\Lambda$ in settings where $D>N$.  

We will show in the next section that MLN linear models are one example of a larger class of models which we term marginally latent matrix-t process (LTP) models. We will show that for MLN linear models this canonical form is characterized by marginalizing over $\Lambda$ and $\Sigma$ so as to replace Equations \eqref{eq:mnllinear3}-\eqref{eq:mnllinear5} with a stochastic process we term a matrix-t process. In Section \ref{section:inference} we build efficient inference methods for MLN linear models by developing tools for inference of Marginally LTP models. 
\section{Modeling Overview} \label{section:modeling}
In this section we will introduce marginally latent matrix-t process (Marginally LTP) models as a flexible class of models capable of describing a wide variety of linear regression, non-linear regression, and time-series models. To build Marginally LTP models we first describe matrix-t processes, then build latent matrix-t processes (LTP), and then generalize this class to Marginally LTP models. 

\subsection{Matrix-Normal and Matrix-T, Distributions and Processes} \label{section:distribution_theory}
To build the class of Marginally LTP models we first review matrix-normal distributions and processes as well as matrix-t distributions and processes highlighting properties we will make use of in this article.

\paragraph{Matrix-Normal Distribution} The matrix-normal distribution is a generalization of the multivariate normal distribution to random matrices. We describe a random $m\times n$ matrix $X$ as being distributed matrix-normal $Y\sim N(M, U, V)$ if $\vec(Y) \sim N(\vec(M), V \otimes U)$ where $U$ is a $m\times m$ covariance matrix and $V$ is a $n \times n$ covariance matrix. 

\paragraph{Matrix-Normal Process} We define a stochastic process $\mathsf{Y}$ as a matrix-normal process on the set $\mathcal{X} = \mathcal{X}^{(1)} \times \mathcal{X}^{(2)}$ and denoted $\mathsf{Y} \sim \mathsf{GP}(\mathsf{M}, \mathsf{K}, \mathsf{A})$ if $\mathsf{Y}$ evaluated on any two finite subsets $\mathbf{x}^{(1)}=(x_1^{(1)}, \dots, x_P^{(1)}) \in \mathcal{X}^{(1)}$ and $\mathbf{x}^{(2)}=(x_1^{(2)}, \dots, x_N^{(2)})\in\mathcal{X}^{(2)}$ is distributed as $Y\sim N(M, K, A)$ where $M_{ij} = \mathsf{M}(x^{(1)}_i, x^{(2)}_j)$, $K_{ij} = \mathsf{K}(x^{(1)}_i, x^{(1)}_j)$, $A_{ij} = \mathsf{A}(x^{(2)}_i, x^{(2)}_j)$ for matrix function $\mathsf{M}$ and kernel functions $\mathsf{K}$ and $\mathsf{A}$. The requirement that  $\mathsf{K}$ and $\mathsf{A}$ be kernel functions implies that the matrices $K$ and $A$ are covariance matricies (\textit{i.e.}, they are symmetric positive definite).

\paragraph{Matrix-t Distribution}  The matrix-t distribution is a generalization of the multivariate-t distribution to random matrices. Like the multivariate-t, the matrix-t can be defined constructively through its relationship to the matrix-normal and inverse Wishart distributions. Let $\Sigma$ denote a random covariance matrix such that  $\Sigma\sim IW(\Xi, \upsilon)$ where $\Xi$ represents a positive semi-definite scale matrix and $\upsilon > 0$. Also suppose that $X\sim N(0, I, V)$. If $CC^T=\Sigma$ then the distribution of $Y=CX$ is denoted as matrix-t such that $Y\sim T(\upsilon, 0, \Xi, V)$. For a random matrix $\eta\sim T(\upsilon, B, K, A)$ the log density of $\eta$ may be written
\begin{multline}
    \log T_{P\times N}(\eta \mid \upsilon, B, K, A)  = \log\Gamma_P\left(\frac{\upsilon+N+P-1}{2}\right) -  \log\Gamma_P\left(\frac{\upsilon+P-1}{2}\right) -\frac{NP}{2}\log\pi \\
    -\frac{N}{2}\log|K| - \frac{p}{2} \log|A| 
    -\frac{\upsilon + N + P - 1}{2} \log \left| I_{p} + K^{-1}[\eta-B]A^{-1}[\eta - B]^T \right| \label{matt_density}
\end{multline}
where $\Gamma_a(b)$ refers to the multivariate gamma function. These results follows directly from \citet[p. 134]{Gupta2018}. 

\paragraph{Matrix-t Process} Through analogy to our definition of matrix normal processes, we define a matrix-t process through its relationship to the matrix-t distribution. We define a stochastic process $\mathsf{Y} \sim \mathsf{TP}(\upsilon, \mathsf{B}, \mathsf{K}, \mathsf{A})$ defined on the set $\mathcal{X} = \mathcal{X}^{(1)} \times \mathcal{X}^{(2)}$ as a matrix-t process if $\mathsf{Y}$ evaluated on any two finite subsets $\mathbf{x}^{(1)}=(x_1^{(1)}, \dots, x_P{(1)}) \in \mathcal{X}^{(1)}$ and $\mathbf{x}^{(2)}=(x_1^{(2)}, \dots, x_N^{(2)})\in\mathcal{X}^{(2)}$ is distributed as $Y\sim T(\upsilon, B, K, A)$ where $\upsilon$ is a scalar strictly greater than zero, $B_{ij}=\mathsf{B}(x^{(1)}_i, x^{(2)}_j)$, $K_{ij} = \mathsf{K}(x^{(1)}_i, x^{(1)}_j)$, and $A_{ij}=\mathsf{A}(x^{(2)}_i, x^{(2)}_j)$ for matrix function $\mathsf{B}$, and kernel functions $\mathsf{K}$ and $\mathsf{A}$. Matrix-t processes can be alternatively seen as a multivariate generalization of Student-t processes which have found widespread use in statistical analysis \citep{Jylanki2011, Shah2014}.

\subsection{Latent Matrix-t Processes (LTPs)}
To allow us to generalize matrix-t processes to a more flexible set of data types, for example count data, we now define LTPs as a generalization of a matrix-t processes. In essence, the following definition generalizes matrix-t processes by defining a stochastic process $\mathsf{Y}$ as a hierarchical process formed by a process $\mathsf{F}$ having parameters that, with appropriate transformation $\phi$, follow a matrix-t process. LTPs allows us to flexibly model complex patterns of covariation in a wide variety of data types. For example, with LTPs we can develop multinomial models with a more flexible covariance structure than the multinomial-Dirichlet, which has strong independence assumptions between multinomial categories \citep{aitchison1980}. The final change compared to our definition of matrix-t processes, is that we now explicitly denote dependence on model hyperparameters which we collectively refer to as $\delta$.  

\begin{definition}{\textbf{Latent Matrix-t Process}} We define a stochastic process $\mathsf{Y}$ as a latent matrix-t process $\mathsf{Y} \sim \mathsf{LTP}(\mathsf{F}, \phi, \upsilon,\mathsf{B}, \mathsf{K}, \mathsf{A}, \delta)$ on the set $\mathcal{X} = \mathcal{X}^{(1)} \times \mathcal{X}^{(2)}$ if $\mathsf{Y}$ evaluated on any $P$ dimensional finite subset $\textbf{x}^{(1)}\in \mathcal{X}^{(1)}$ and any $N$ dimensional finite subset $\textbf{x}^{(2)}\in \mathcal{X}^{(2)}$ is distributed 
\begin{align}
    Y &\sim f(\pi, \delta) \\
    \pi &= \phi^{-1}(\eta) \\
    \eta &\sim T(\upsilon, B(\delta), K(\delta), A(\delta)). 
\end{align}
where $\eta$ denotes a $P \times N$ real valued matrix, $B(\delta)$ a $P\times N$ dimensional real valued matrix function of parameters $\delta$ defined by $[B(\delta)]_{ij} = \mathsf{B}(x^{(1)}_i, x^{(2)}_j, \delta)$, $K(\delta)$ is a $P\times P$ covariance matrix defined as $[K(\delta)]_{ij} = \mathsf{K}(x^{(1)}_i, x^{(1)}_j, \delta)$,  $A(\delta)$ is an $N\times N$ covariance matrix defined as $[A(\delta)]_{ij} = \mathsf{A}(x^{(2)}_i, x^{(2)}_j, \delta)$, $\upsilon$ is a scalar subject to $\upsilon>0$, $\pi$ is an element of a space $\Pi$ defined via the one-to-one mapping $\phi^{-1}:\mathcal{R}^{P\times N} \rightarrow \Pi$, $f$ denotes a likelihood function with parameters $\pi$ and $\delta$ which is itself an evaluation of the process $\mathsf{F}$ evaluated on a finite subset of the set $\Pi$. 
\end{definition}

Before proceeding further we motivate the generality of the above definition with two examples.

\subsubsection{LTPs for analysis of Generalized Linear Models} 
Our first example demonstrates how LTPs can be used to infer Bayesian generalized linear models. 
Consider the following univariate exponential generalized linear model (GLM) for observed data $Y$ and covariates $X$
\begin{align*}
    Y_j &\sim \text{Exponential}(\lambda_j) \\
    \lambda_j &= -\frac{1}{\eta X_{\cdot j}} \\
    \eta &\sim N(0, \Sigma) \\
    \Sigma &\sim IW(\Xi, u). 
\end{align*}
Noting that $\eta = N(0, \Sigma)$ can alternatively be written as $\eta \sim N(0, \Sigma, 1)$ and noting the definition of the matrix-t, we can state $\eta$ marginalizing over $\Sigma$ as $\eta \sim T(u, 0, \Xi, 1)$. Thus, marginalizing over $\Sigma$ in the above model results in an LTP with parameters $f = \prod_{i=1}^N \text{Exponential}(Y_i \mid \lambda_i), \upsilon=u, B = 0, K = \Xi, A = 1, \delta = \emptyset$ where $\emptyset$ denotes the empty set and  $\phi^{-1}$ is defined by the relation $\lambda_j = [\phi^{-1}(\eta)]_j = -\frac{1}{\eta X_{\cdot j}}$. This result demonstrates that inference of $\eta$ in the above univariate exponential GLM can instead be replaced by inference of an LTP with the above parameters.   
 
\subsubsection{LTPs for Non-Linear Modeling of Microbiome Time-Series}
Motivated by the non-linear time-series model in \citet{aijo2016}, our second example illustrates how LTPs can be used to construct a non-linear multinomial time-series model for the analysis of microbiome data.
Let $Y$ denote a $D\times T$ matrix of counts such that $Y_{\cdot j}$ represents the sequence count vector of $D$ taxa measured in the sample from the $t$-th time-point. This data could be modeled as 
\begin{align*}
    Y_{\cdot t} &\sim \text{Multinomial}(\pi_{\cdot t}) \\
    \pi_{\cdot t} &=\text{ALR}^{-1}(\eta_{\cdot t}) \\
    \eta &\sim T(\upsilon, B, K, A) \\
    B_{\cdot t} &= 0_{d-1} \\
    K_{i,j} &= \kappa^2\exp(-\gamma^{-2}[d_\text{phylo}(i,j)]^2) \\
    A_{t,s} &= \alpha^2 \exp ( -\rho ^{-2} (t-s)^2)
\end{align*}
where $A_{t,s}$ denotes a radial basis function kernel describing the similarity between time-points $t$ and $s$ and $d_\text{phylo}(i,j)$ represents the phylogenetic (evolutionary) distance between microbial log-ratios $i$ and $j$ as introduced in \citet{silverman2017}. Intuitively, this specification models a microbial time-series as a composition evolving through time subject to non-linear dynamics and with the prior assumption that evolutionary more similar taxa are likely to behave more similarly. Additionally, the term $B_{\cdot t} = 0_{D-1}$ can be interpreted as saying we have little prior knowledge regarding which taxa are the most abundant at any time-point\footnote{When all ALR coordinates are equal to zero this corresponds to $\pi_i=1/D$ for all taxa $i$.}. Additionally the relationship between the matrix-t, the inverse Wishart and the matrix-normal implies that this model can also be seen as a multinomial, log-ratio transformed matrix-normal (\textit{i.e.}, matrix-variate version of a logistic-normal) with uncertainty in the covariance matricies of the matrix-normal described by an inverse Wishart distribution. In this way the specification of $K$ and $A$ is likely less sensitive to miss-specification than if we had $\eta \sim N(B, K, A)$. In particular, this implies that the scalar $\upsilon$ represents our uncertainty in our specification of $K$ and $A$. 

Finally, this non-linear time-series model can be seen to represent an LTP with the following specifications: $f = \prod_{t=1}^T \text{Multinomial}(Y_{\cdot t} \mid \pi_{\cdot t}), 
    \phi^{-1} = ALR^{-1},
    \upsilon = \upsilon,
    B_{\cdot t} = 0_{D-1},
    K_{i,j} = \kappa^2\exp(-\gamma^{-2}[d_\text{phylo}(i,j)]^2),$
    $A_{t,s} = \alpha^2 \exp ( -\rho ^{-2} (t-s)^2),
    \delta = \{ \kappa, \gamma, \alpha, \rho \} $
which together also imply that $\mathbf{x}^{(1)} = \{1, \dots, D-1\} \in \mathcal{X}^{(1)}$, $\mathbf{x}^{(2)} = \{1, \dots, T\} \in \mathcal{X}^{(2)}$, and correspondingly that $P = D-1$ and $N=T$.

\subsection{Marginally LTP Models}
We next generalize our definition of LTPs to a larger class which we term Marginally LTP models. This definition is straight forward, we define Marginally LTP models as those models which have a marginal that is an LTP. 

\begin{definition}{\textbf{Marginally LTP models}} If a model described by the joint distribution $p(\eta, \Psi, Y)$ may be written as $p(\Psi \mid \eta, Y) \, p(\eta,Y)$  where $p(\eta,Y)$ is a latent matrix-\textit{t} process, we refer to $p(\eta, \Psi, Y)$ as a \textit{Marginally LTP} model and $p(\eta, Y)$ as the model's collapsed representation. 
\end{definition}

In the next two subsections we demonstrate that Marginally LTP models provide a rich class of models. We give two examples of Marginally LTP models, the first represents a generalization of the multinomial logistic normal linear models introduced in Section \ref{section:multinomial}, the second represents a flexible class of models for inference in multivariate non-Gaussian time-series. 

\subsubsection{Generalized Multivariate Conjugate Linear (GMCL) Models} \label{section:GMCL}
As in Section \ref{section:multinomial}, let us consider $Y$ to represent $N$ independent $D$-variate measurements and consider $X$ to represent $N$ sets of $Q$-dimensional covariates. We define generalized multivariate conjugate linear (GMCL) models as 
\begin{align}
    Y_{\cdot j} &\sim f(\pi_{\cdot j})  \label{mod_e1}\\
    \pi_{\cdot j} &= \phi^{-1}(\eta_{\cdot j})\label{mod_e2} \\
    \eta_{\cdot j} &\sim N(\Lambda X_{\cdot j}, \Sigma) \label{mod_e3}\\
    \Lambda &\sim N(\Theta, \Sigma, \Gamma) \label{mod_e4}\\
    \Sigma &\sim IW(\Xi, \upsilon) \label{mod_e5}. 
\end{align}
We may describe the joint density of this model as $p(\Lambda, \Sigma, \eta, Y)$ which can be factored as $p(\Lambda, \Sigma \mid \eta, Y) \, p(\eta, Y)$. Therefore, in parallel to the definition of Marginally LTP models we may equate $\Psi = \{\Lambda, \Sigma\}$.  In Appendix \ref{section:collapsedform} we show that $p(\eta, Y)$ is LTP matrix-t with parameters
\begin{align*}
    B &= \Theta X\\
    K &= \Xi \\
    A&=I_N+X^T\Gamma X
\end{align*}
and where parameters $f$, $\phi$, and $\upsilon$ are defined flexibly as in Equations \eqref{mod_e1} and \eqref{mod_e2}. Finally this model specification implies that $\{\Theta, \Gamma, \Xi\} \in \delta$. This result demonstrates that all GMCL models are Marginally LTP models. Further, by letting $f$ denote the multinomial distribution and $\phi^{-1}$ denote the inverse ALR transform, we see that the multinomial logistic-normal linear models introduced in Section \ref{section:multinomial} are a special case of GMCL models. 

\subsubsection{Generalized Multivariate Dynamic Linear Models (GMDLMs)}
 
We develop a flexible class of multivariate time-series models for non-Gaussian observations. We term this class of models generalized multivariate dynamic linear models (GMDLMs). GMDLMs represent an extension of the multivariate dynamic linear models introduced in \citet{Quintana1987} and developed further in \citet[Ch. 16]{west1997} to non-Gaussian observations. Using the notation from \citet[Ch. 16]{west1997} we let $\eta_t^T$ denote a row-vector (\textit{i.e.}, the transpose of the $t$-th column of $\eta$). We define the GMDLM as
\begin{align}
    Y_{\cdot j} &\sim f(\pi_{\cdot j})  \label{eq:dlm1}\\
    \pi_{\cdot j} &= \phi^{-1}(\eta_{\cdot j}) \label{eq:dlm2}\\ 
    \eta_t^T &= F_t^T\Theta_t + \nu_t^T, \quad \nu_t \sim N(0, \gamma_t \Sigma) \label{eq:dlm3} \\
    \Theta_t &= G_t \Theta_{t-1} + \Omega_t, \quad \Omega_t \sim N(0, W_t, \Sigma) \label{eq:dlm4} \\
    \Theta_0 &\sim N(M_0, C_0, \Sigma)  \label{eq:dlm5}\\
    \Sigma &\sim IW(\Xi, \upsilon) \label{eq:dlm6}
\end{align}
where $\Theta_t$ represents a $Q \times P$ matrix describing the state of the time-series at time $t$, $G_t$ denotes the $Q\times Q$ state transition matrix at time $t$, $F_t$ denotes a $1\times Q$ vector describing a linear model relating the latent space to the parameters $\eta_t$,  $\Sigma$ is a $P \times P$ covariance matrix specifying the covariation between the $P$ dimensions of the time-series, $W_t$ is a $Q\times Q$ covariance matrix describing the covariation of the perturbations affecting latent states, and $\gamma_t$ is a  scalar allowing an analyst to weight the importance of select observations ($\gamma_t$ is typically equal to $1$).  

The joint model for the GMDLM can be written $p(\Theta, \Sigma, \eta, Y)$ which can be factored as $p(\Theta, \Sigma \mid \eta, Y) \, p(\eta, Y)$. To parallel the definition of Marginally LTP models, here we have $\Psi = \{\Theta, \Sigma\}$. In Appendix \ref{section:dlm} we demonstrate that $p(\eta, Y)$ is an LTP with parameters 
\begin{align*}
    B &= \begin{bmatrix} 
      \vertbar & &\vertbar &&\vertbar \\
      \alpha_1 &\cdots &\alpha_t &\cdots &\alpha_T \\
      \vertbar & &\vertbar & &\vertbar 
      \end{bmatrix}  \\
    \alpha_t &= (F_t^T\mathcal{G}_{t:1}M_0)^T \\
    K &= \Xi \\
    A_{t,t-k} &= 
      \begin{cases}
        \gamma_t + F_t^T\left[
        W_t + \sum_{\ell=t}^{2}\mathcal{G}_{t:\ell}W_{\ell-1}\mathcal{G}^T_{\ell:t} + \mathcal{G}_{t:1}C_0\mathcal{G}_{1:t}^T
        \right]F_t \text{ if } k=0\\
        F_t^T\left[\mathcal{G}_{t:t-k+1}  W_{t-k} +
    \sum_{\ell=t-k}^2 \mathcal{G}_{t:\ell} W_{\ell-1}G^T_{\ell:t-k}
    + \mathcal{G}_{t:1} C_0G^T_{1:t-k} \right]F_{t-k} \text{ if } k>0
      \end{cases}
\end{align*}
where we have introduced $\mathcal{G}_{t:\ell}$ as a short hand notation for the product $G_t\cdots G_\ell$. Additionally here LTP parameters $\upsilon$, $f$, and $\phi$ are defined flexibly as in Equations \eqref{eq:dlm1} and \eqref{eq:dlm2} and we have hyperparameters $\{ \Xi, M_0, C_0, W_1, \dots, W_T, \gamma_1, \dots, \gamma_T, G_1, \dots, G_T, F_1,\dots, F_T \}\in \delta$. This result demonstrates that GMDLMs are a special case of Marginally LTP models. 

\section{Inference in Marginally LTP Models} \label{section:inference}
Our overarching goal was to develop efficient and accurate posterior inference for multinomial logistic-normal linear models which are a special case of Marginally LTP models. In this section we demonstrate how the canonical LTP form of Marginally LTP Models can be exploited for efficient inference of this larger model class. We consider two types of parameters, $\eta$ which are distributed matrix-t and $\Psi$ which are marginalized out of a model to produce a LTP form. The inference scheme we introduce will involve two steps: 1) sampling $\eta$ conditioned on observed data ($p(\eta \mid Y)$), 2) sampling $\Psi$ conditioned on $\eta$ and observed data ($p(\Psi \mid \eta,  Y)$). In Section \ref{section:cu} we introduce this sampling scheme which we refer to as the collapse-uncollapse (CU) sampler. In Section \ref{section:laplace} we further build on the CU sampler by introducing a Laplace approximation as a means of sampling $p(\eta \vert Y)$. In Sections \ref{section:inferece_conditional} we discuss efficient means of sampling $p(\Psi \mid \eta , Y)$ for the GMCL model and GMDLM models introduced in the previous section.  

\subsection{The Collapse-Uncollapse (CU) Sampler} \label{section:cu}
Consider the task of sampling from the posterior distribution of a Marginally LTP model with joint density $p(\Psi, \eta, Y)$. The corresponding posterior density can be decomposed as
\[p(\eta, \Psi \mid Y) = p(\Psi \mid \eta, Y)\frac{p(\eta, Y)}{p(Y)}. \]
This decomposition implies the following: given a model described by the joint distribution $p(\eta, \Psi, Y)$, we may sample from the posterior by first sampling from the posterior of the collapsed (LTP) model $p(\eta,Y)$ and then given that sample of $\eta$ and the observed $Y$ we may then sample $\Psi$ from the conditional $p(\Psi \mid \eta, Y)$. Together the sample of $\eta$ and $\Psi$ then represents a single sample from the posterior of the Marginally LTP model, $p(\Psi, \eta \mid Y)$ (Algorithm \ref{algorithm:cu}). 

\begin{algorithm}[H] \label{algorithm:cu}
\caption{The Collapse-Uncollapse (CU) Sampler for Marginally LTP Models}
\KwData{$Y, \upsilon, B, K, A$}
\KwResult{$S$ samples of the form $\{\Psi^{(s)}, \eta^{(s)}\}$}
Sample $\{\eta^{(1)},\dots,\eta^{(S)}\}\sim p(\eta \mid Y)$ where $p(\eta \mid Y)$ is an LTP\;
\ForPar{s in $\{1,\dots,S\}$}{
Sample $\Psi^{(s)}\sim p(\Psi \mid \eta^{(s)}, Y)$\;
}
\end{algorithm}

The CU Sampler for Marginally LTP Models therefore requires two features for effcient inference. First, we require an efficient means of producing samples from the collapsed (LTP) form $p(\eta \mid Y)$. As we will show in Section \ref{section:simulations}, sampling $p(\eta \mid Y)$ can be more efficient than sampling $p(\Psi, \eta \mid Y)$ just by virtue of the fact that the former has fewer dimensions. Therefore the CU sampler alone can be more efficient than sampling the full (uncollapsed) model. Still, in Section \ref{section:laplace} we develop a Laplace approximation for $p(\eta \mid Y)$ which can further improve efficiency. Second, we require an efficient means of sampling from the posterior conditional $p(\Psi \mid \eta, Y)$. In Section \ref{section:inferece_conditional} we discuss efficient means of sampling $p(\Psi \mid \eta, Y)$ for the GMCL and GMDLM models.  

Our rationale for focusing on the CU sampler for inference in Marginally LTP models is as follows. We expect that many Marginally LTP models (such as those introduced in Section \ref{section:modeling}) have partial conjugacy. Methods such as Metropolis-within-Gibbs have been popular methods of exploiting this conjugacy \citep{cargnoni1997}. Yet, by embedding MCMC steps within a Gibbs sampler techniques such as adaptation \citep{Gelman2015} or approximate methods such as Laplace approximations may not make sense. In contrast, the CU sampler allows the non-conjugate sampling to occur up front (in the sampling of $p(\eta \mid Y)$) so that such techniques can be used. Moreover, after multiple samples of $\eta$ have been produced, uncollapsing the model can be done in parallel for each sample of $\eta$. Therefore, the CU sampler may be advantageous as it permits the use of adaptive or approximate methods for sampling the non-conjugate model components and permits a degree of parallelism not allowed by Metropolis-within-Gibbs.

\subsection{Laplace Approximation for the Collapsed Form} \label{section:laplace}
Sampling $p(\eta \mid Y)$ is often the major computational bottleneck when inferring Marginally LTP models via the CU sampler. To accelerate inference we developed a Laplace approximation for the density $p(\eta \mid Y)$. This approximation is defined as $q(\eta \mid Y) = N(\vec\, \hat{\eta}, H^{-1}(\vec\,\hat{\eta}))$ where $\hat{\eta}$ denotes the MAP estimate of  $p(\eta \mid Y)$ and $H^{-1}(\vec\,\hat{\eta})$ denotes the inverse Hessian matrix of $\log p(\eta \mid Y)$ evaluated at the point $\vec\,\hat{\eta}$ where $\hat{\eta}$ represents the \textit{maximum a posteriori} (MAP) estimate. That is $\hat{\eta}$ is defined as the solution to the following optimization problem
\begin{equation} \label{eq:optimgoal}
\hat{\eta} = \argmin_{\eta\in\mathcal{R}^{P\times N}}\left[-\log p(\eta \mid Y)\right]. 
\end{equation}

We hypothesized that such a Laplace approximation would provide an accurate approximation to an LTP posterior based on the following observations. First, the matrix-normal can provide a good approximation for the matrix-t for suitably large $\upsilon$ \citep[p. 137]{Gupta2018} as it is both globally symmetric and log-convex about the MAP estimate. Second, all exponential family likelihoods are log-convex with respect to their natural parameters \citep{Jordan2010}. Notably, the ALR parameters represent the natural parameters of the multinomial distribution. Together these features suggested that the posterior of a LTP could be approximated as globally symmetric and log-convex like the multivariate normal. 

Developing an efficient Laplace approximation for LTP models required closed form solutions for the gradient and Hessian of LTPs. To develop these tools note that by Bayes rule we may write 
\begin{align}
     -\log p(\eta \mid Y) &\propto - \log f(Y \mid \phi^{-1}(\eta))  + p(\eta) \nonumber \label{collapsed_density}.  
\end{align}
By linearity of the derivative operator we may write the gradient and Hessian of $-\log p(\eta \mid Y)$ as
\begin{align*}
    -\frac{d\log p(\eta \mid Y)}{d\vec(\eta)} &= -\frac{d \log f(Y \mid  \phi^{-1}(\eta))}{d\vec(\eta)}- \frac{d\log p(\eta)}{d\vec(\eta)} \\
    -\frac{d^2\log p(\eta|Y)}{d\vec(\eta)d\vec(\eta)^T} &= -\frac{d^2 \log f(Y \mid \phi^{-1}(\eta))}{d\vec(\eta)d\vec(\eta)^T}- \frac{d^2\log p(\eta)}{d\vec(\eta)d\vec(\eta)^T}.
\end{align*}
Thus we find that calculating the gradient and Hessian of LTPs reduces to calculating the gradient and hessian of $\log f(Y \mid \phi^{-1}(\eta))$ and the matrix-t density $\log p(\eta \mid X)$ separately. The additive structure of the gradient and Hessian are central to generalizing this approach to a variety of different observation distributions $f$ and transformations $\phi^{-1}$. In Appendix \ref{section:ghmatt} we provide the gradient and Hessian for the matrix-t density. With these results, to derive a flexible class of multinomial logistic-normal models we only need to provide the gradient and Hessian for the logit-parameterized multinomial which we state in Appendix \ref{section:app_multinomgh}. 

\subsection{Efficient Sampling from Posterior Conditionals} \label{section:inferece_conditional}
The second step of the CU sampler involves sampling from the density $p(\Psi \mid \eta, Y)$. While the density of $p(\Psi \mid \eta, Y)$ is specific to the particular Marginally LTP model, we develop efficient means of sampling from this density for the GMCL and GMDLM models in Appendices \ref{section:collapsedform} and \ref{section:dlm} respectively. In particular, for both the GMCL and GMDLM models we make use of the fact that $\Psi$ is conditionally independent of $Y$ given $\eta$, that is $p(\Psi \mid \eta , Y) = p(\Psi \mid \eta)$. This conditional independence also reduces sampling from the conditionals to computing the posterior distribution of standard Bayesian multivariate linear regression for the GMCL model and conjugate multivariate dynamic linear models for the GMDLM model. That is, for both the GMCL and GMDLM models sampling the conditionals reduces to posterior inference for equivalent Bayesian Gaussian models that have been well described previously. 

\subsection{Software for Multinomial Logistic-Normal Linear Models} \label{section:software}
For inference of multinomial logistic normal linear models we developed the R package \textit{stray} \citep{Silverman2019}. \textit{Stray} implements the CU sampler with Laplace approximation described above using optimized C++ code. Estimation of $\hat{\eta}$ is performed using the L-BFGS optimizer which we have found provides efficient and stable numerical results. 

Additionally all code required to reproduce the results of the next two sections, including the alternative implementations of multinomial logistic-normal linear models discussed \ref{section:simulations} is available as a GitHub repository at https://github.com/jsilve24/stray\_paper\_code. 
\section{Simulations} \label{section:simulations}
We performed a series of simulation studies to evaluate both the CU sampler and the Laplace Approximation in terms of accuracy and efficiency of posterior inference in multinomial logistic-normal linear models. To evaluate the utility of the collapse-uncollapse sampler we compared Hamiltonian Monte Carlo (HMC) of the full model (HMC Uncollapsed) to the CU sampler where sampling of the collapsed (LTP) form was performed using HMC (HMC Collapsed). Both HMC implementations were inferred using the highly optimized  No-U-Turn-Sampler provided in the Stan modeling language \citep{Gelman2015} which has been frequently for the analysis of MLN models \citep{aijo2016, Silverman2018}. To further evaluate the utility of the Laplace approximation to the collapsed form in the CU sampler (LA Collapsed), we used the \textit{stray} software package described in Section \ref{section:software}. Finally, to compare LA Collapsed to an alternative scheme for approximate inference, we included two mean-field automatic-differentiation Variational Bayes (VB) implementations \citep{Kucukelbir2015}. The first was a VB approximation to the full form (VB Uncollapsed), the second was a VB approximation to just the collapsed form of the CU sampler (VB Collapsed). As VB Uncollapsed was unstable in practice, often resulting in error during optimization, only the results form VB Collapsed could be shown below. 

\begin{figure}
    \centering 
    \includegraphics[width=5.5in]{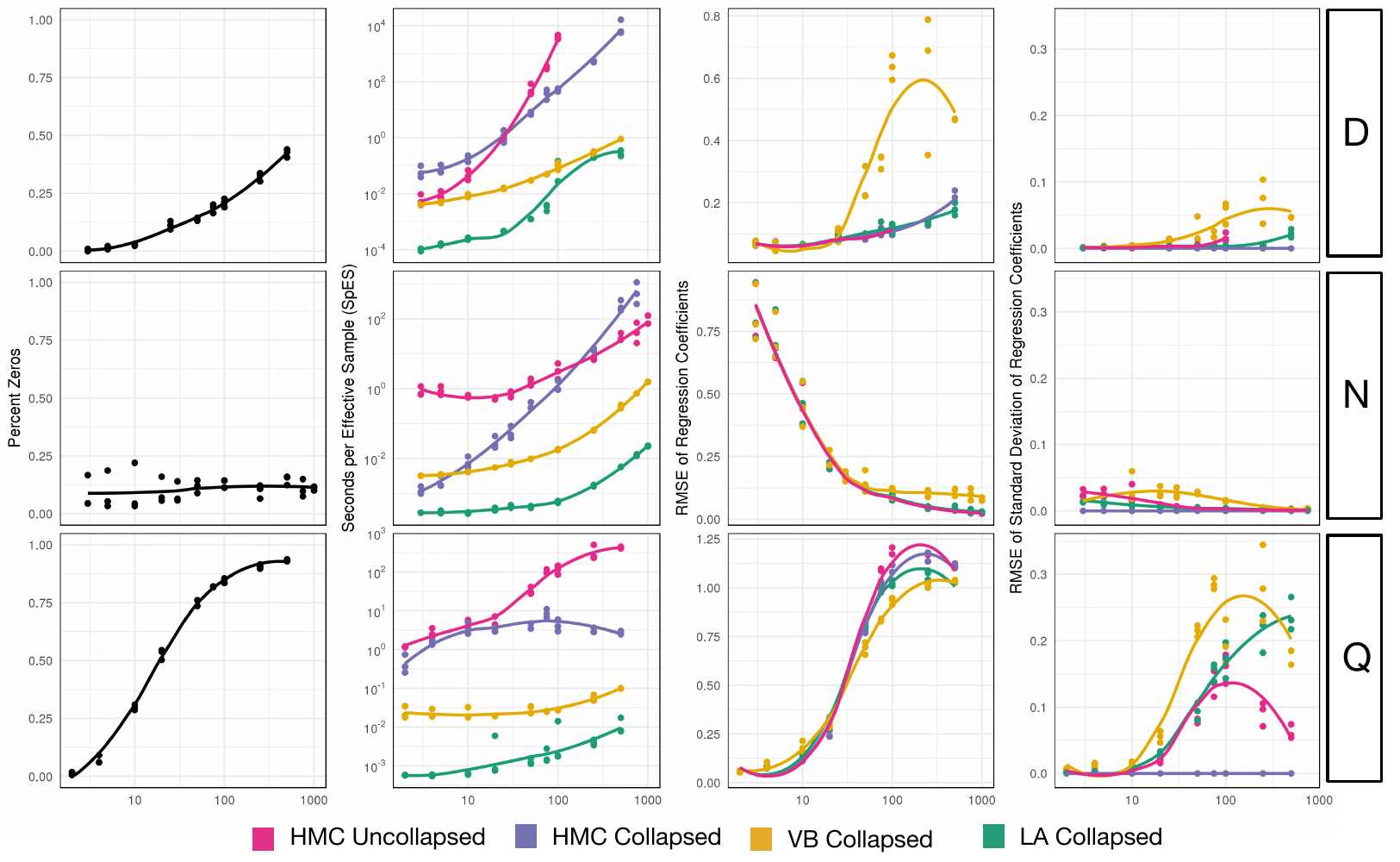}
    \caption[Simulation study of different multinomial logistic normal linear model implementations.]{\textbf{Simulation study comparing multinomial logistic normal linear model implementations.} Each row of plots depicts simulation results for varying a different simulation parameter ($D$, the number of multinomial categories; $N$, the number of samples; and $Q$, the number of covariates). The percent of counts that were zero is given in the left column. Implementations were compared in terms of efficiency (measured by SpES), accuracy of point estimation (measured by RMSE of Regression Coefficients), and accuracy of uncertainty quantification (measured by RMSE of Standard Deviation of Regression Coefficients).}
    \label{fig:simulations_main}
\end{figure}

In order to compare these implementations, we created a series of simulations based on the likelihood model described in Equations \eqref{eq:mnllinear1}-\eqref{eq:mnllinear3}. We identified three key parameters,  the sample size ($N$),  the observation dimension ($D$), and the number of model covariates ($Q$) which we varied in order to test each implementation over a wide variety of conditions. We choose the tuple $\{N=100, D=30, Q=5\}$ as our base condition and independently varied each simulation parameter from that base condition ($N$ from 10 to 1000, $D$ from 3 to 500, and $Q$ from 2 to 500). Importantly, since we hypothesized that the Laplace and variational approximations would be challenged at high levels of zero values, we developed these simulations such that the proportion of zero counts also increased with increasing $D$ and $Q$ (Figure \ref{fig:simulations_main}). Additionally, to account for the stochastic nature of the simulations, three simulations were performed for each tuple $\{N, D, Q\}$. For each simulation, each of the five implementations were fit and allowed a maximum of 48 hours to produce 2000 samples. Further details of the simulation and model fitting procedure can be found in Appendix \ref{section:app_simulation}. 

To quantify the accuracy and efficiency of each implementation we defined the following performance metrics. As a measure of efficiency, we calculated the average number of seconds needed for the implementation to produce one independent sample from the target posterior (\textit{i.e.}, Seconds per Effective Sample - SpES). Specifying independent samples is important as HMC samplers produce autocorrelated samples. To quantify the accuracy of point estimates from each implementation (\textit{i.e.}, either the posterior mean or MAP) we defined the root mean squared error of the point estimate for $\Lambda$ from its true simulated value.  Notably, given finite $N$ we do not expect that any implementation will be able to perfectly reconstruct the true simulated value for $\Lambda$; rather, this metric provides a means of comparing the relative performance of each implementation. Finally, to quantify the accuracy of uncertainty quantification from each implementation we compared posterior intervals against those of the HMC Collapsed model which was taken as a gold standard. In particular, we define the root mean squared error of standard deviations as the average difference between the estimated posterior standard deviations, $sd(\Lambda_{ij})$, compared to the estimates produced by HMC Collapsed.

Overall, we hypothesized that LA Collapsed would be more efficient than the other implementations but would have low accuracy of uncertainty quantification when the percent of zeros in the observed data was higher than 30\%. The later portion of this hypothesis is based on the observation that all exponential family distributions parameterized by their natural parameters (\textit{e.g.}, a multinomial parameterized by log-ratio coordinates) is globally log-convex \citep{Jordan2010} whereas the matrix-t distribution is only log-convex near the mean. We therefore hypothesized that a weaker likelihood would lead the Laplace approximation to have thinner tails than the true posterior. In practice LA Collapses provided nearly identical estimates of posterior uncertainty to both HMC implementations up to over 90\% data sparsity. Additionally, LA Collapsed provided nearly identical point estimates to both HMC implementation over the full spectrum of simulations. Finally, LA Collapsed was often up to 5 orders of magnitude faster than HMC and often 1-2 orders of magnitude faster than VB. 

\subsection{Computational Efficiency}

Overall, the CU sampler with a Laplace approximation (LA Collapsed) provides the most efficient inference across all tested conditions. More specifically LA Collapsed displays speed ups of between 1 to 5 orders of magnitude in comparison to HMC Collapsed and Uncollapsed and often between 1-2 order of magnitude compared to VB Collapsed. Notably, HMC Uncollapsed fails to complete sampling within 48 hours for $D>100$. 

Beyond the high efficiency of LA Collapsed, our results also demonstrate that the CU sampler can improve inference in HMC without the use of approximate inference methods. These results likely stem from the smaller number of dimensions in HMC for the collapsed versus uncollapsed implementations. Most noticeably, the collapsed representation completely removes dependency on $Q$ from HMC run-times as $\Lambda$ is marginalized out of the collapsed representation. However, for large $N$ the HMC Uncollapsed is more efficient than HMC Collapsed. This result may reflect that the heavy tails of the matrix-t distribution produce a more challenging geometry for HMC than the expanded  matrix normal and inverse Wishart forms. Such a result has been well described previously for both univariate and multivariate-t distributions \citep[Section 20]{StanUM}.

\subsection{Point Estimation}

Overall point estimation using LA Collapsed (\textit{i.e.}, MAP estimates) is nearly identical to point estimation using either HMC Collapsed or HMC Uncollapsed (\textit{i.e.}, mean estimates). In contrast point estimation using VB Collapsed can produce substantially larger errors, especially for large values of $D$. Overall these results demonstrate that the CU sampler maintains accuracy in point estimation and that MAP estimation provides an excellent approximation to the mean in multinomial logistic normal linear models.

\subsection{Uncertainty Quantification}
Beyond accuracy of point estimates, we also wanted to study the accuracy of estimates of uncertainty from each implementation. We consider the HMC Collapsed implementation to be the gold standard which we base our metric \textit{RMSE of standard deviations} on. Except for values of $Q$ greater or equal to 250 (where the proportion of zero values is $>$90\%), the uncertainty estimates of LA Collapsed are identical to those of both HMC implementations. Yet, at larger values of $Q$, when sparsity is $>$90\%, we see differences not only between LA Collapsed and HMC but between the two HMC implementations themselves. There are two possible explanations for this. First, that stray has a slightly better point estimation accuracy in these same large $Q$ simulations could point to the fact that stray is correct and instead HMC estimates of uncertainty are incorrect due to the often small effective sample size for large $Q$. Alternatively, this could support our previous hypothesis that the Laplace approximation has higher error in uncertainty quantification with higher data sparsity. Given the ergodicity of HMC it seems more likely that the Laplace approximation is in error in these regions of high sparsity. Yet, that the approximation only begins to show error when sparsity is $>$90\% is notable. Beyond LA Collapsed and the HMC implementations, VB Collapsed consistently demonstrates higher error in uncertainty quantification as compared to the other implementations. 

Finally, to provide context regarding the size of the differences in uncertainty quantification, we provide direct visualizations of posterior intervals for all four implementations in Figure \ref{fig:successcase} and \ref{fig:failcase}. These two simulations were chosen to highlight a case in which LA Collapsed is highly accurate (\ref{fig:successcase}) in terms of uncertainty quantification and a case in which it differs from HMC estimates (\ref{fig:failcase}).  Notably, visualization of posterior intervals consistently demonstrated that the posterior mean was centered symmetrically in the 95\% credible regions. This symmetry suggested that our metric \textit{RMSE of standard deviations} captures much of the discrepancies in uncertainty quantification without higher order moments. Additionally, for context we include a fifth implementation, PCLM (pseudo-count augmented linear model). The PCLM uses a pseudo-count based estimate of $\eta$ which ignores the multinomial count variation. Such approximations are common in the analysis of microbiome sequence count data \citep{silverman2017,Gloor2016}. Unsurprisingly, this PCLM implementation demonstrates substantially higher error rates than any of the other implementations (Figure \ref{fig:pclm}).


\section{Real Data Analysis}
Crohn's Disease (CD) is a type of inflammatory bowel disease that has been linked to aberrant immune response to intestinal microbiota \citep{jostins2012, khor2012, gevers2014}. To demonstrate that LA Collapsed (from the R package \textit{stray}) provides an accurate and efficient means of modeling real microbiome data, we reanalyzed a previously published study comparing microbial composition in the terminal ileum of subjects with CD to healthy controls. Only LA Collapsed could efficiently scale to this data size (49 taxa, 250 samples, 4 covariates). To allow us to compare to alternative implementations we randomly subset the data to contain 83 samples. On this subset HMC Uncollapsed and VB Collapsed repeatedly failed to run due to numerical instability. In addition, LA Collapsed produced posterior estimates nearly identical to HMC Collapsed but more than 1000 times faster. 

\begin{figure}[h]
    \centering
    \includegraphics[width=\textwidth]{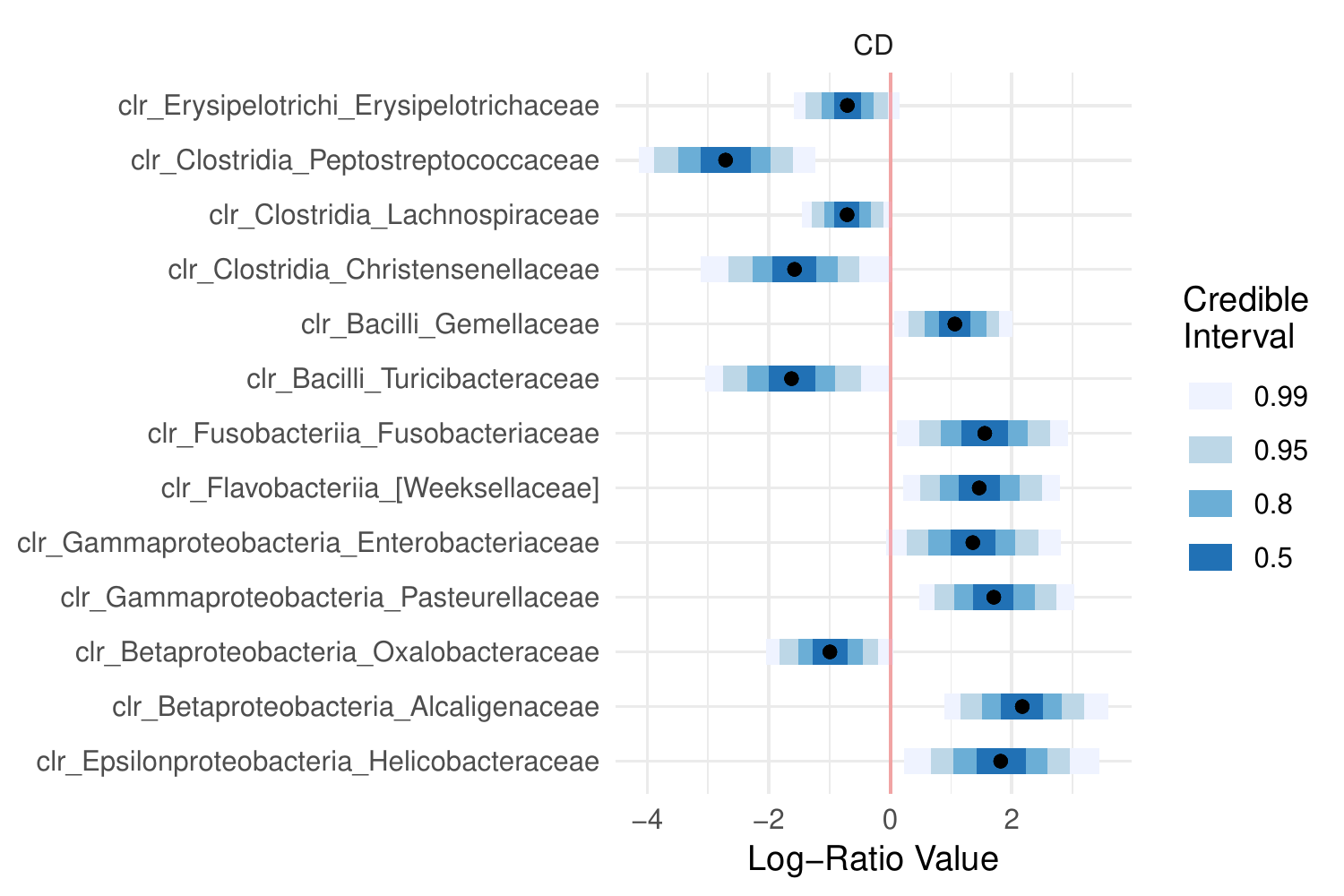}
    \caption[Results of \textbf{stray} analysis of Real Data]{\textbf{Posterior mean and credible intervals for $\Lambda$ of \textit{stray} (LA Collapsed) applied to real data.} Only the 12 families found to be associated with Crohn's Disease (CD) (\textit{i.e.}, Posterior 95\% credible region not covering zero) are shown. Taxa are denoted as clr\_[class]\_[family]. $\Lambda$ is represented in centered log-ratio (CLR) coordinates rather than additive log-ratio (ALR) so that each coordinate could be identified with a different bacterial genera.}
    \label{fig:realdata}
\end{figure}

Using the four model implementations introduced in Section \ref{section:simulations}, a Bayesian multinomial logistic normal linear model was fit to investigate the relationship between bacterial composition and CD. For both the full data-set and the subset, our regression model was defined for the $j$-th sample by the covariate vector \[x_j=[1,x_{j(\text{CD})},x_{j(\text{Inflamed})},x_{j(\text{Age})}]^T\] where  $x_{j(\text{CD})}$ is a binary variable denoting whether the $j$-th sample was from a patient with CD or a healthy control, $x_{j(\text{Inflamed})}$ a binary variable denoting inflammation at time of sample collection, $x_{j(\text{Age})}$ denoting age of the subject, and the preceding 1 represents a constant intercept. A detailed description of our prior assumptions is given in Appendix \ref{section:app_priors} and results of posterior predictive checks are shown in Figure \ref{fig:ppc}.

Even though all four implementations were initialized identically, both the HMC Uncollapsed and VB Collapsed implementations repeatedly resulted in errors due to numerical instability. Thus only LA Collapsed and the HMC Collapsed implementations could fit this model for even the subset dataset. Whereas the HMC Collapsed model took approximately 30 minutes, LA Collapsed took only 3 seconds. Thus LA Collapsed is over 1000 times faster than HMC Collapsed on real data.  Additionally, posterior estimates of $\Lambda$ produced by both the HMC Collapsed and LA Collapsed implementations are nearly identical (Figure \ref{fig:real_data_comparision}). These results demonstrate that in real data scenarios LA Collapsed can provide efficient and accurate posterior inference. 

By modeling the full dataset we found the centered log-ratio (CLR) coordinates corresponding to 12 genera to be associated with CD status (95\% credible interval not covering 0; Figure \ref{fig:realdata}). These results are in general agreement with prior analyses \citep{gevers2014}. As in prior analyses, we find a substantial increase in the abundance of proteobacteria in CD versus healthy controls. Similarly, we find that the families Pasteurellaceae and Enterobacteriacaeae, Gemellaceaem, and Fusobacteriaceae are highly enriched and that the class Clostridia are depleted in CD. Notably, Fusobacteria has been independently suggested as a marker of IBD \citep{strauss2011, kostic2012}. These findings serve to validate our results and build confidence in our methods. 

In contrast, our results differ from prior analyses of this data in certain respects. We find that the family Peptostreptococcaceae is likely decreased in CD versus healthy controls and we find no association for Veillonellaceae. Three factors support our results. First, our analysis accounts for  count variation and compositional constraints whereas prior analyses have not. Notably, the handing of count variation and compositional constraints can have substantial impact on conclusions in the analysis of sequence count data \citep{mcmurdie2014, silverman2018naught, gloor2017}. Second, Peptostreptococcaceae has been found to be decreased in CD based on the analysis of independent data \citep{imhann2018}. Third, in visualizing the count data for Peptostreptococcaceae and Veillonellaceae (Figure \ref{fig:compare_to_prior_analysis}) we find no visual difference in Veillonellaceae but a notable difference in Peptostreptococcaceae. Therefore, we conclude that our approach has revealed novel associations in this data and excluded potentially spurious conclusions.
\section{Conclusion}

In this work we have developed efficient inference methods for the analysis of multinomial logistic-normal linear models through the use of a marginally latent matrix-t Process (LTP) representation. Through the use of the CU sampler and Laplace approximation (referred to as LA Collapsed) we have demonstrated up to 5 orders of magnitude increases in computational efficiency compared to HMC while preserving accuracy of point estimation and uncertainty quantification.  Beyond the inference of multinomial logistic-normal linear models, we find that our approach generalizes to a larger class of models including non-linear regression (LTP) models and time-series models (GMDLMs). Importantly, the key computational advance we have introduced relates to the marginal representation of multinomial logistic-normal linear models which is identical to that of multinomial logistic-normal GMDLMs. For this reason we hypothesize that the LA Collapsed sampler should share similar gains in efficiency for these other multinomial logistic-normal models. Yet, the performance of our Laplace approximation under observation distributions beyond the log-ratio parameterized multinomial is more uncertain. Notably, we hypothesize that our results could generalize to other exponential family distributions parameterized by natural parameters due to the fact that such distributions are globally log-concave. Yet, we expect that there are other observation distributions where a Laplace approximation to the LTP form may be sub-optimal. Rather than resorting to MCMC, we suggest that methods of particle refinement of the initial Laplace approximation (\textit{e.g.}, parallel MCMC steps for each sample from the LTP form, or sequential importance resampling) may be more efficient. We believe such extensions are prime areas for future work. 

Until this point we have not considered the presence of unknown hyperparameters in the LTP form (\textit{i.e.} we have considered $\delta$ as given). Yet for a number of Marginally LTP models we expect estimation of such hyperparameters will be of interest. For example, within the GMDLM model we anticipate researchers may want to allow the terms $W_t$ to be subject to their own stochastic model. This would in turn require that some portion of $\delta$ is unknown. Unfortunately, the multitude of ways in which such unknown hyper-parameters may arise is beyond the scope of this work. However, we note that our early efforts at extending multinomial logistic-normal linear models to handle variance components with unknown scale suggests that such hyperparameters can be learned efficiently during MAP estimation of $\hat{\eta}$. Alternatively, when the number of unknown parameters is small, such parameters could be choosen by cross-validation as is often done for choosing unknown parameters in kernel functions when using Gaussian process models. Still, we believe that future work incorporating unknown hyper-parameters into the LTP form as required by specific Marginally LTP models would be impactful. 
\section*{Acknowledgements}

 JDS was supported in part by the Duke University Medical Scientist Training Program (GM007171). JDS and LAD were supported in part by the Global Probiotics Council, a Searle Scholars Award, the Hartwell Foundation, an Alfred P. Sloan Research Fellowship, the Translational Research Institute through Cooperative Agreement NNX16AO69A, the Damon Runyon Cancer Research Foundation, the Hartwell Foundation, and NIH 1R01DK116187-01. SM and KR would like to acknowledge the support of grants NSF IIS-1546331, NSF DMS-1418261, NSF IIS-1320357, NSF DMS-1045153, NSF DMS1613261, and NSF DEB-1840223. The authors thank Rachel Silverman, Liat Shenhav, and Shaobo Han for their helpful comments. 

\newpage
\bibliographystyle{jasa}
\bibliography{main}

\newpage
\appendix
\section{Generalized Conjugate Multivariate Linear (GMCL) Models} \label{section:collapsedform}
Here we demonstrate that the GMCL models defined in Equations \eqref{mod_e1}-\eqref{mod_e5} are Marginally LTP models by deriving their collapsed (LTP) form. Additionally, we demonstrate that uncollapsing the LTP form can be done efficiently. 

\subsection{Derivation of Collapsed Form}
Our goal in this section is to derive a collapsed (or marginalized) form of the GMCL model defined in Equations \eqref{mod_e1}-\eqref{mod_e5} that will have a posterior distribution of the form $p(\eta \mid Y,X)$. To do this, note that Equations \eqref{mod_e3}-\eqref{mod_e5} can alternatively be written as 
\begin{align}
    \eta &= \Lambda X+E^\eta \quad E^\eta \sim N(0, \Sigma, I_N) \label{acol_e1}\\
    \Lambda &= \Theta + E^\Lambda \quad E^\Lambda \sim  N(0, \Sigma, \Gamma) \label{acol_e2}\\
    \Sigma &\sim IW(\Xi, \upsilon).\label{acol_e3}
\end{align}

We will first marginalize $\Lambda$ in the above model by using the following affine transformation properties of the matrix normal distribution. Given an $m \times m$ matrix $A$, a $p \times m$ matrix $B$, and $n \times r$ matrix $C$, and a random matrix $X\sim N(M, U, V)$; then for a random matrix $Z = A + BXC$ we have $Z \sim N(A+BMC, BUB^T, C^TVC)$ \citep[p. 64]{Gupta2018}. We note that Equations \eqref{acol_e1} and \eqref{acol_e2} can be collapsed into the following form
\begin{align}
    \eta &= \Theta X + E^\Lambda X + E^\eta \quad E^\eta \sim N(0, \Sigma, I_N) \quad E^\Lambda \sim  N(0, \Sigma, \Gamma) \nonumber \\
    &= \Theta X + E^* \quad E^\star \sim N(0, \Sigma, I_N + X^T \Gamma X).  \label{acol_e4}
\end{align}
Thus we may rewrite Equations \eqref{acol_e1}-\eqref{acol_e3} as 
\begin{align}
    \eta &= \Theta X + E^* \quad E^\star \sim N(0, \Sigma, I_N + X^T \Gamma X) \label{acol_e5}\\ 
    \Sigma &\sim IW(\Xi, \upsilon).\label{acol_e6}
\end{align}
By using the definition of the matrix-t given in Section \ref{section:distribution_theory} we can marginalize over $\Sigma$ in Equations \eqref{acol_e5} and \eqref{acol_e6} to get 
\[ \eta \sim T(\upsilon, \Theta X, \Xi, I_N+X^T\Gamma X). \]
Finally, incorporating equations \eqref{mod_e1} and \eqref{mod_e2} allow us to write the marginalized form of GMCL models, $p(\eta \mid Y, X)$, as
\begin{align*}
    Y &\sim f(\pi)  \\
    \pi &= \phi^{-1}(\eta) \\
    \eta &\sim T(\upsilon, B, K, A)
\end{align*}
where $B=\Theta X$, $K=\Xi$, and $A=I_N + X^T\Gamma X$. 

\subsection{Efficient Form for Uncollapsing}
Here we demonstrate that for GMCL models, the conditional posterior $p(\Lambda, \Sigma \vert \eta, Y, X)$ can be computed and sampled efficiently. As $\Lambda$ and $\Sigma$ are conditionally independent of $Y$ given $\eta$ in GMCL models, we may write 
\begin{equation*}
p(\Lambda, \Sigma \mid \eta, Y, X) = p(\Lambda, \Sigma \mid \eta, X) = p(\Lambda \mid \Sigma, \eta, X)\, p(\Sigma \mid \eta, X).
\end{equation*} 
The right hand side of the above equation represents the posterior of a multivariate conjugate linear model that can be sampled efficiently using the following relations \citep[p. 32]{Rossi2012}:
\begin{align*}
    \upsilon_N &= \upsilon+N \\
    \Gamma_N &= (XX^T+\Gamma^{-1})^{-1} \\
    \Lambda_N &= (\eta X^T+\Theta\Gamma^{-1})\Gamma_N \\
    \Xi_N &= \Xi + (\eta - \Lambda_N X)(\eta - \Lambda_N X)^T + (\Lambda_N - \Theta)\Gamma^{-1}(\Lambda_N - \Theta)^T \\
    p(\Sigma | \eta, X) &= IW(\Xi_N, \upsilon_N)\\
    p(\Lambda | \Sigma, \eta, X) &= N(\Lambda_N, \Sigma, \Gamma_N).
\end{align*}
\section{Generalized Multivariate Dynamic Linear Model (GMDLM)} \label{section:dlm}
Here we demonstrate that the GMDLM defined in Equations \eqref{eq:dlm1}-\eqref{eq:dlm6} is a marginally LTP model and show how the theory and tools we develop improves inference in this model class. Additionally we provide a recursive procedure for uncollapsing an LTP to a GMDLM (\textit{i.e.}, an efficient recursive approach to sampling $p(\Theta, \Sigma \mid \eta, Y)$. 

\subsection{Derivation of Collapsed Form}
Here we derive the marginal distribution of $\eta$ in terms of quantities $F_t$, $G_t$, $W_t$, $\Sigma$, $M_0$ and $C_0$. As all densities involved are multivariate or matrix-variate normal, the result must also be multivariate or matrix-variate normal and thus fully described by the mean and covariance of $\eta$. To derive the mean and covariance we first derive a useful alternative representation of $\eta^T_t$ with respect to $\Theta_{t-k-1}$ for some positive integer $k < t$. 

Substituting Equation \eqref{eq:dlm4} into Equation \eqref{eq:dlm3} allows $\eta_t^T$ be expressed with respect to $\Theta_{t-1}$ as 
\begin{equation}
    \eta^T_t = F_t^T G_t \Theta_{t-1} + F_t^T \Omega_t + \nu_t^T. \label{eq:dlm_deriv0} 
\end{equation}
Repeated substitution of $\Theta_{t-k}$ leads to the following form for $\eta_t^T$ in terms of $\Theta_{t-k-1}$  
\begin{equation}
    \eta_t^T = F_t^T\mathcal{G}_{t:t-k}\Theta_{t-k-1} + F_t^T\Omega_t + \sum_{\ell=t}^{t-k-1} F_t^T\mathcal{G}_{t:\ell} \Omega_{\ell-1} + \nu_t^T \label{eq:dlm_deriv1}
\end{equation}
where $\mathcal{G}_{t:t-k}$ is shorthand for $G_tG_{t-1}\cdots G_{t-k}$.

Note that if $X\sim N(0, I, I)$, $AA^T = U$ and $BB^T=V$ then for $Z = M+ AXB^T$ we have the marginal distribution of $Z$ (\textit{e.g.} marginalizing over $X$) given by $Z \sim N(M, U, V)$. This identity in combination with \eqref{eq:dlm5} allows us to marginalize over the random variables ${\Omega_t, \dots, \Omega_1, \nu_t}$ in Equation \eqref{eq:dlm_deriv1} giving
\begin{equation}
    \eta^T_t \sim N\left(F_t^T\mathcal{G}_{t:1}M_0, \gamma_t + F_t^T\left[
    W_t + \sum_{\ell=t}^{2}\mathcal{G}_{t:\ell}W_{\ell-1}\mathcal{G}^T_{\ell:t} + \mathcal{G}_{t:1}C_0\mathcal{G}_{1:t}^T
    \right]F_t, \Sigma\right) \label{eq:dlm_deriv2}.
\end{equation}

Next we calculate $Cov(\eta_t^T, \eta_{t-k}^T)$. In parallel to Equation \eqref{eq:dlm_deriv0} we may write $\eta_{t-k}^T$ as 
\begin{equation}
    \eta_{t-k}^T = F_{t-k}^TG_{t-k}\Theta_{t-k-1} + F_{t-k}^T\Omega_{t-k} + \nu_{t-k}^T. \label{eq:dlm_deriv3}
\end{equation}
Using Equation \eqref{eq:dlm_deriv1} and \eqref{eq:dlm_deriv3} along with the fact $Cov(AX_1 + BX_2, Y) = A \, Cov(X_1,Y) + B \, Cov(X_2, Y)$ and that $Cov(\Theta_s, \nu_\ell) = Cov(\Theta_s, \Omega_\ell) = Cov(\Omega_\ell, \nu_s)=0$ for all $s$ and $\ell$, we can write
\begin{align}
Cov(\eta_t^T, \eta_{t-k}^T) &= F_t^T\mathcal{G}_{t:t-k}Var(\theta_{t-k-1})G_{t-k}^TF_{t-k} + F_t^T\mathcal{G}_{t:t-k+1}Var(\Omega_{t-k})F_{t-k} \label{eq:dlm_deriv4}
\end{align}
where $Var(\Theta_{t-k-1})$ can be written recursively as 
\begin{equation*}
    Var(\Theta_{t-k-1}) = G_{t-k-1}Var(\Theta_{t-k-2})G_{t-k-1}^T + Var(\Omega_{t-k-1})
\end{equation*}
where $Var(\Omega_{t-k-1}) = \Sigma\otimes W_{t-k-1}$.  Combining this recursive form with equation \eqref{eq:dlm_deriv4} gives 
\begin{multline}
        Cov(\eta_t^T, \eta_{t-k}^T) = F_t^T\mathcal{G}_{t:t-k+1} (\Sigma \otimes W_{t-k})F_{t-k} +
    \sum_{\ell=t-k}^2 F_t^T\mathcal{G}_{t:\ell}(\Sigma \otimes W_{\ell-1})G^T_{\ell:t-k}F_{t-k} \\
    + F_t^T\mathcal{G}_{t:1}(\Sigma \otimes C_0)G^T_{1:t-k}F_{t-k}. \label{eq:dlm_deriv5}
\end{multline}

Together Equations \eqref{eq:dlm_deriv2} and \eqref{eq:dlm_deriv5} characterize the marginal distribution of $\eta_t^T$ in terms of $F_t$, $G_t$, $W_t$, $\Sigma$, $M_0$ and $C_0$. Noting that if $X\sim N(M, U, V)$ then $X^T\sim N(M^T, V, U)$,  if follows that
\begin{align*}
    \eta &\sim N(B, \Sigma, A) \\
    B &= \begin{bmatrix} 
      \vertbar & &\vertbar &&\vertbar \\
      \alpha_1 &\cdots &\alpha_t &\cdots &\alpha_T \\
      \vertbar & &\vertbar & &\vertbar 
      \end{bmatrix}  \\
    \alpha_t &= (F_t^T\mathcal{G}_{t:1}M_0)^T \\
    A_{t,t-k} &= 
      \begin{cases}
        \gamma_t + F_t^T\left[
        W_t + \sum_{\ell=t}^{2}\mathcal{G}_{t:\ell}W_{\ell-1}\mathcal{G}^T_{\ell:t} + \mathcal{G}_{t:1}C_0\mathcal{G}_{1:t}^T
        \right]F_t \text{ if } k=0\\
        F_t^T\left[\mathcal{G}_{t:t-k+1}  W_{t-k} +
    \sum_{\ell=t-k}^2 \mathcal{G}_{t:\ell} W_{\ell-1}G^T_{\ell:t-k}
    + \mathcal{G}_{t:1} C_0G^T_{1:t-k} \right]F_{t-k} \text{ if } k>0
      \end{cases}
\end{align*}
Finally, using the  marginalization property of the matrix normal and the inverse Wishart used in our definition of the matrix-t distribution and incorporating Equations \eqref{eq:dlm1}, \eqref{eq:dlm2} and \eqref{eq:dlm6} it follows
\begin{align*}
    Y &\sim f(\pi)\\
    \pi &= \phi^{-1}(\eta) \\
    \eta &\sim T(\upsilon, B, \Xi, A). 
\end{align*}
This result demonstrates that the GMDLM model is a marginally LTP model.

\subsection{Efficient Form for Uncollapsing}
Here we provide an efficient means of sampling from the conditional density $p(\Theta, \Sigma \mid \eta, Y)$ for the GMDLM. First we recognize that $\Theta$ is conditionally independent of $Y$ given $\eta$. Therefore, our task simplifies to sampling from $p(\Theta, \Sigma \mid \eta)$. The problem is identical to the standard filtering and simulation smoothing problem solved by \citet[p. 603-604]{west1997}. Again, the problem is defined by the following model (which we will refer to as the MDLM model) 
\begin{align}
    \eta_t^T &= F_t^T\Theta_t + \nu_t^T, \quad \nu_t \sim N(0, \gamma_t \Sigma) \\
    \Theta_t &= G_t \Theta_{t-1} + \Omega_t, \quad \Omega_t \sim N(0, W_t, \Sigma) \\
    \Theta_0 &\sim N(M_0, C_0, \Sigma)\\
    \Sigma &\sim IW(\Xi, \upsilon).    
\end{align}

Following \citet{west1997}, below we restate the filtering and retrospective recursions needed to sample from $p(\Theta, \Sigma \mid \eta)$. Note that all densities in this subsection are conditional on the parameters $F_t$, $G_t$, $W_t$, $\Sigma$, $M_0$ and $C_0$ but that this dependence has been suppressed for notational simplicity. Let us introduce $\upsilon_t$ and $\Xi_t$ as filtering parameters at step $t$. Further, we define $\upsilon_0=\upsilon$ and $\Xi_0=\Xi$. As a final piece of notation we introduce $H^T_{t}$ as a shorthand for the set $\{\eta^T_t, \dots, \eta^T_1\}$

\subsubsection{Filtering Recursions for MDLM Model}
(1) Posterior at $t-1$:
\begin{align*}
    p(\Sigma \mid H^T_{t-1}) &\sim IW(\Xi_{t-1}, \upsilon_{t-1}) \\
    p(\Theta_{t-1} \mid  \Sigma, H^T_{t-1}) &\sim N(M_{t-1}, C_{t-1}, \Sigma)
\end{align*}
(2) Prior at $t$:
\begin{align*}
    A_t &= G_tM_{t-1} \\
    R_t &= G_tC_{t-1}G_t^T+W_t \\
    p(\Sigma \mid H^T_{t-1}) &\sim IW(\Xi_{t-1}, \upsilon_{t-1}) \\
    p(\Theta_t \mid  \Sigma, H^T_{t-1}) &\sim N(A_{t}, R_{t}, \Sigma)
\end{align*}
(3) One-step ahead forecast at $t$:
\begin{align*}
    f_t^T &= F_t^TA_t \\
    q_t &= \gamma_t + F_t^TR_tF_t \\
    p(\Sigma \mid H^T_{t-1}) &\sim IW(\Xi_{t-1}, \upsilon_{t-1}) \\
    p(\eta_t \mid  \Sigma, H^T_{t-1}) &\sim N(f_t, q_{t}\Sigma)
\end{align*}
(4) Posterior at $t$:
\begin{align}
    e_t^T &= \eta_t^T-f_t^T  \nonumber \\
    S_t &= \frac{R_tF_t}{q_t} \nonumber \\
    M_t &= A_t + S_te_t^T \nonumber\\ 
    C_t &= R_t - q_tS_tS_t^T \nonumber\\
    \upsilon_t &= \upsilon_{t-1} + 1 \nonumber\\
    \Xi_t &= \Xi_{t-1} + \frac{e_t e_t^T}{q_t} \label{eq:filter} \\
    p(\Sigma \mid H^T_{t-1}) &\sim IW(\Xi_{t}, \upsilon_{t}) \nonumber\\
    p(\Theta_t \mid  \Sigma, H^T_t) &\sim N(m_t, C_{t}, \Sigma) \nonumber
\end{align}
Equation \eqref{eq:filter} differs slightly from the presentation in \citet{west1997} as the parameterization of the inverse-Wishart we employ throughout this paper differs from that source. Throughout this work we use the following parmeterization for a random matrix $\Sigma \sim IW(\Xi, \upsilon)$:
\begin{align*}
    p(\Sigma) \propto \left| \Sigma \right|^{-(P + \upsilon + 1)/2} \exp \Big( -\frac{1}{2} \textrm{tr} \big( \Xi \Sigma^{-1} \big) \Big).  
\end{align*}

\subsubsection{Simulation Smoothing Recursion}
The recursions provided here follow directly from \citet[p. 268]{Prado2010}\\
(1) Sample $\Sigma \sim IW(\Xi_T, \upsilon_T)$ and then $\Theta_T \sim N(M_t, C_t, \Sigma)$. \\
(2) For each time $t$ from $T-1$ to $0$, sample $p(\Theta_{t} | \Theta_{t+1}, H_T^T) \sim N(M_t^*, C_t^*, \Sigma)$ where 
\begin{align*}
Z_t&=C_tG_{t+1}^TR_{t+1}^{-1}\\
M_t^* &= M_t + Z_t(\Theta_{t+1}-a_{t+1}) \\
C_t^* &= C_t - Z_t R_{t+1}Z_t^T.
\end{align*}

\section{Gradient and Hessian Calculations for the Matrix-T Distribution} \label{section:ghmatt}
Here we are concerned with calculating the gradient and Hessian of  
\[\log p(\eta \mid Y) \propto -\frac{\upsilon + N + P-1}{2} \log \left| I_{P} + K^{-1}(\eta - B)A^{-1}(\eta-B)^T\right|.\] 
Letting $S = I_{P} + K^{-1}(\eta - B)A^{-1}(\eta-B)^T$ we concern ourselves with calculating the quantities $\frac{d \log |S|}{d\vec(\eta^T)}$ and $\frac{d \log |S|}{\vec(d\eta)\vec(d\eta)^T}$. We will use the following identity from matrix calculus  \citep[pg. 1]{Minka2000}: 
\begin{align*}
d\log |S| &= Tr(S^{-1} dS)  \\
dS &= d(I_{P} + K^{-1}(\eta - B)A^{-1}(\eta-B)^T) \\
   &= d(K^{-1}(\eta A^{-1} \eta^T  -\eta A^{-1} B^T - B A^{-1} \eta^T)) \\
   & = K^{-1}(d\eta A^{-1}\eta^T + \eta A^{-1} d\eta^T - d\eta A^{-1}B^T-B A^{-1} d\eta^T) \\
   & = K^{-1}(d\eta(A^{-1}\eta^T-A^{-1}B^T) + (\eta A^{-1} - B A^{-1})d\eta^T) \\
   & = K^{-1}(d\eta C + C^T d\eta^T)
\end{align*}
where in the last line we have defined the $N \times P$ matrix $C = A^{-1}(\eta^T - B^T)$. Further simplifying and using the identities $Tr(A) = Tr(A^T)$ and $Tr(AB)= Tr(BA)$ for matrices $A$ and $B$ we get 
\begin{align}
d\log|S| &= Tr(S^{-1}K^{-1}(d\eta C + C^T d\eta^T)) \nonumber \\
&= Tr(S^{-1}K^{-1}d\eta C) + Tr(S^{-1}K^{-1}C^T d\eta^T) \nonumber \\
&= Tr(CS^{-1} K^{-1} d\eta) + Tr(CK^{-1}S^{-T}d\eta) \nonumber \\
&= Tr(C(S^{-1}K^{-1} + K^{-1}S^{-T})d\eta) \nonumber \\
&=\vec([C(S^{-1}K^{-1}+K^{-1}S^{-T})]^T)^T\vec(d\eta) \nonumber \\
d\log|S|&=\vec((S^{-1}K^{-1}+K^{-1}S^{-T})C^T)^T\vec(d\eta) \label{eq:grad1differential} \\
\frac{d\log|S|}{\vec(d\eta)} &= \vec((S^{-1}K^{-1}+K^{-1}S^{-T})C^T)^T  \nonumber
\end{align}

The Hessian $H=\frac{d^2\log|S|}{\vec(d\eta)\vec(d\eta)^T}$ can then be calculated from equation \eqref{eq:grad1differential} by taking the differential again and manipulating the result into the following canonical form $d^2\log|S| = \vec(d\eta)^TH\vec(d\eta)$. In particular we make use of the following identities $\vec(ABC) = (C^T\otimes A)\vec(B)$ and $d(S^{-1}) = -S^{-1}dS S^{-1}$. We also make use of the vec-transposition matrix defined by $T_{m,n}\vec(A) = \vec(A^T)$ where $A$ is an $m\times n$ matrix and $T_{m,n}$ is an $mn\times mn$ permutation matrix. The vec-transposition matrix also satisfies the following properties $T_{m,n} = T^T_{n,m} = T^{-1}_{n,m}$. 
\begin{align*}
d^2\log|S| &= \vec((S^{-1}K^{-1}+K^{-1}S^{-T})dC^T+d(S^{-1})K^{-1}C^T + K^{-1}d(S^{-T})C^T)^T\vec(d\eta) \\
&= \left[\vec((S^{-1}K^{-1}+K^{-1}S^{-T})dC^T)^T+\vec(d(S^{-1})K^{-1}C^T)^T + \vec(K^{-1}d(S^{-T})C^T)^T\right]\vec(d\eta) \\
&= \left[\#1 + \#2 + \#3 \right]\vec(d\eta) \\
\#1 & = \vec((S^{-1}K^{-1}+K^{-1}S^{-T})d\eta A^{-1})^T\\ 
& = ((A^{-1}\otimes(S^{-1}K^{-1}+K^{-1}S^{-T})) \vec(d\eta))^T \\
& = \vec(d\eta)^T(A^{-1}\otimes (S^{-1}K^{-1}+K^{-1}S^{-T}))^T \\ 
\#2 &= -\vec(S^{-1}dSS^{-1}K^{-1}C^T)^T \\
&=-((CK^{-1}S^{-T}\otimes S^{-1})\vec(dS))^T \\
&=-\vec(dS)^T(S^{-1}K^{-1}C^T \otimes S^{-T}) \\
\vec(dS)^T &= \vec(K^{-1}(d\eta C + C^T d\eta^T))^T \\
&=\vec(K^{-1}d\eta C)^T + \vec(K^{-1}C^Td\eta^T)^T \\
&=((C^T\otimes K^{-1})\vec(d\eta))^T + ((I_{D-1}\otimes K^{-1}C^T)\vec(d\eta^T))^T \\
&=\vec(d\eta)^T(C\otimes K^{-1}) + \vec(d\eta^T)^T(I_P\otimes CK^{-1}) \\
\#2&= [-\vec(d\eta)^T(C\otimes K^{-1}) - \vec(d\eta^T)^T(I_P\otimes CK^{-1})](S^{-1}K^{-1}C^T\otimes S^{-T}) \\
&= -\vec(d\eta)^T(CS^{-1}K^{-1}C^T\otimes K^{-1} S^{-T}) - \vec(d\eta^T)^T(S^{-1}K^{-1}C^T\otimes CK^{-1} S^{-T}) \\
&= -\vec(d\eta)^T(CS^{-1}K^{-1}C^T\otimes K^{-1} S^{-T}) - \vec(d\eta)^TT_{N, P}(S^{-1}K^{-1}C^T\otimes CK^{-1} S^{-T}) \\
\#3 &= \vec(K^{-1}d(S^{-T})C^T)^T\\ 
&=-\vec(K^{-1}S^{-T}dS^TS^{-T}C^T) \\
&=-((CS^{-1}\otimes K^{-1}S^{-T})\vec(dS^T))^T \\
&=-\vec(dS^T)^T(S^{-T}C^T\otimes S^{-1}K^{-1}) \\
\vec(dS^T)^T&= \vec((d\eta C + C^T d\eta^T)^TK^{-1})^T \\
&= ((K^{-1}C^T\otimes I_P)\vec(d\eta))^T + ((K^{-1}\otimes C^T)\vec(d\eta^T))^T \\
\#3 &= [-\vec(d\eta)^T(CK^{-1}\otimes I_P)-\vec(d\eta^T)^T(K^{-1}\otimes C)](S^{-T}C^T \otimes S^{-1}K^{-1}) \\
&= -\vec(d\eta)^T(CK^{-1}S^{-T}C^T\otimes S^{-1}K^{-1})-\vec(d\eta)^TT_{N,D-1}(K^{-1}S^{-T}C^T\otimes CS^{-1}K^{-1}) \\
d^2\log|S| &= \vec(d\eta)^T [(A^{-1}\otimes (S^{-1}K^{-1} + K^{-1}S^{-T}))^T - (CS^{-1}K^{-1}C^T\otimes K^{-1}S^{-T})  \\
& \qquad -(CK^{-1}S^{-T}C^T) \otimes S^{-1}K^{-1})  \\
& \qquad -T_{N, D-1}((S^{-1}K^{-1}C^T\otimes CK^{-1}S^{-T})+(K^{-1}S^{-T}C^T\otimes CS^{-1}K^{-1}))] \vec(d\eta) 
\end{align*}
Summarizing the above results together we obtain
\begin{align*}
S &= I_{P} + K^{-1}(\eta-B)A^{-1}(\eta-B)^T \\
C &= A^{-1}(\eta -B)^T\\
R &= S^{-1}K^{-1}\\
\frac{d\log |S|}{\vec(d\eta)} &= \vec((R+R^T)C^T)^T \\
L &= (CRC^T\otimes R^T) \\
\frac{d^2\log|S|}{\vec(d\eta)\vec(d\eta)^T} &= (A^{-1}\otimes (R+R^T)) - (L + L^T) - T_{N, D-1}[(RC^T\otimes CR^T)+(R^TC^T\otimes C R)].
\end{align*}
The following computational trick makes evaluation of this Hessian far more computationally efficient. We may quickly calculate $T_{m,n}X=X^*$ for an $m\times m$ matrix $X$ having already computed $X$ by noting that for $i\in{1 \dots m}$ and $j\in{1\dots n}$ we can write $X^*_{(i-1)n+j, \cdot} = X_{(j-1)m+i, \cdot}$ where $X_{l,\cdot}$ denotes the $l$-th row of the matrix $X$. 
\section{Gradients and Hessians for  the Log-Ratio Parameterized Multinomial} \label{section:app_multinomgh}
Unfortunately we cannot provide a general form for the gradient and Hessian of all possible likelihoods $f(Y \mid \phi^{-1}(\eta))$. For the purposes of this article, here we derive the gradient and Hessian for the case where $f$ is multinomial and $\phi^{-1}$ is the inverse ALR transform:
\[\sum_j \log \text{Multinomial}(Y_{\cdot j}  \mid \text{ALR}_D^{-1}(\eta_{\cdot j}))\]
which for notational simplicity we will refer to as $g$. Thus our goal is to find efficient forms for calculating $g$, $\frac{dg}{d\vec(\eta)}$ and $\frac{d^2g}{d\vec(\eta)d\vec(\eta)^T}$. Using the fact that $\log \text{Multinomial}(Y_{\cdot j} \mid \pi_{\cdot j}) \propto Y_{1j}\log \pi_{1j}+\dots +Y_{Dj}\log \pi_{Dj}$ and Equation \eqref{eq:alrdef} we can write 
\begin{equation*}
    g = \sum_{j=1}^N\left(\sum_{i=1}^{D-1}\eta_{ij}Y_{ij}-n_j\log\left(1+\sum_{i=1}^{D-1}e^{\eta_{ij}}\right) \right). 
\end{equation*}
Differentiating with respect to $\eta_{ij}$ gives 
\[ \frac{dg}{d\eta_{ij}} =  Y_{ij} - n_j \frac{e^{\eta_{ij}}}{1+\sum_i e^{\eta_{ij}}}.\]
Differentiating again with respect to $\eta_{k\ell}$ gives
\begin{equation*}
\frac{d^2g}{d\eta_{ij}d\eta_{k\ell}} = 
\begin{cases}
-n_j\left( \frac{e^{\eta_{ij}}}{1+\sum_i e^{\eta_{ij}}} - \frac{e^{2\eta_{ij}}}{(1+\sum_i e^{\eta_{ij}})^2} \right) & \text{if } \ell=j, i=k\\
n_j \left( \frac{e^{\eta_{ij}}e^{\eta_{kj}}}{(1+\sum_i e^{\eta_{ij}})^2} \right) & \text{if } \ell=j, i\neq k\\
0 & \text{if } \ell\neq j. \\
\end{cases}
\end{equation*}
These results directly imply the following matrix forms. 
\begin{align*}
    O &= \exp \eta \\
    m &= 1_N + O^T 1_{D-1}\\
    \rho &= \vec(O)\oslash \vec(1_{D-1}m^T) \\
    n &= 1^T_DY \\
    g &= - \vec(\eta)^T\vec(Y_{/D\cdot}) - n\odot \log(m) \\
    \frac{dg}{d\vec(\eta)} &= (\vec(Y_{/D\cdot}) - \vec(1_Dn)\odot\rho)^T \\
    W^{(j)} &= n_j(\rho_{(j)}\rho_{(j)}^T-\text{diag}(\rho_{(j)})) \\
    \frac{d^2g}{d\vec(\eta)d\vec(\eta)^T} &= \text{diag}\left(W^{(1)}, \dots, W^{(N)}\right)
\end{align*}
where $\exp X$ and $\log X$ refers to the element-wise exponentiation and logarithm of a matrix $X$, $\odot$ and $\oslash$ refer to element-wise product and division respectively, $Y_{/D\cdot}$ refers to the first $D-1$ rows of the matrix $Y$,  $\rho_{(j)}$ denotes elements $(j-1)(D-1)+1$ to $j(D-1)$ in the vector $\rho$, and $\text{diag}(X_1, \dots, X_D)$ refers to a block diagonal matrix where the $i$-th block is $X_i$. 

\section{Accelerated Matrix-T Gradients via Sylvester's Determinant Identity}
Sylvester's determinant identity states that for matrices $A$ and $B$ of size $m\times n$ and $n\times m$ respectively, $|I_m + AB| = |I_n+BA|$. This relationship can be used to speed up calculation of the log-likelihood and gradient of the matrix-$t$ distribution when $N < P$ as the the log determinant or inverse of the matrix $S$ can dominate computational time. To take advantage of this speed up we note that we can replace the relations given in Appendix \ref{section:ghmatt} with 
\begin{align}
    S &= I_N + A^{-1}(\eta - B)^TK^{-1} (\eta - B) \\
    C &= K^{-1}(\eta - B)\\
    R &= S^{-1}A^{-1}\\
    \frac{d\log|S|}{d\Vec{(\eta})} &= \Vec{(C(R+R^T))}^T.
\end{align}
While this result can greatly accelerate inference for matrix-t gradients when $P \gg N$, this result provides only minimal improvement for calculating the corresponding Hessian terms. Therefore we suggest that, for simplicity, the Hessian form provided in Appendix \ref{section:ghmatt} be used even if $P \gg N$. 

\section{Simulations and Model Fitting} \label{section:app_simulation}
To compare the performance of the multiple multinomial logistic-normal linear model implementations described in Section \ref{section:simulations} over a range of sample sizes ($N$), observation dimensions ($D$), and covariate dimensions ($Q$), we created the following simulation scheme. For each evaluated triple ($N$, $D$, $Q$), three simulated data-sets were created based on the multinomial logistic-normal linear model
with the following specified likelihood:
\begin{align*}
    Y_{\cdot j} &= \textrm{ALR}_D^{-1}(\eta_{\cdot j}) \\
    \eta_{\cdot j} &= N(\Lambda X_{\cdot j}, \Sigma) \\
    \Lambda &\sim N(\Theta, \Sigma, \Gamma) \\
    \Sigma &\sim IW(\Xi, \upsilon)
\end{align*}
with $\upsilon=D+10$, $\Xi=I_{D-1}$. Additionally $X$, $\Lambda$, and $\Sigma$ were simulated as 
\begin{align*}
    \Lambda &\sim N(0, I, I) \\
    \Sigma &\sim IW(\Xi, \upsilon) \\
    X &\sim N(0, I, I).
\end{align*}
The percent of zero counts naturally increased with large $D$ or large $Q$ relative to other parameters. We took advantage of this behavior to study the performance of all implementations in sparse data regimes. \\

All implementations were compiled and run using gcc version 6.2.0, R version 3.4.2, and Intel(R) Math Kernel Library version 2019 where possible. All replicates of the simulated count data were supplied to the various implementations independently and the models were fit on identical hardware, allotted 64GB RAM, 4 cores, and restricted to a 48-hour upper limit on run-time.
\section{Priors for Crohn's Disease Data} \label{section:app_priors}
\subsection{Data Preprocessing}
Sequence count data was obtained from the R package MicrobeDS (github.com/twbattaglia/MicrobeDS). Only samples from the terminal illeum from healthy donors and patients with Crohn's Disease, who had no recent history of steroids, antibiotics, or biologics were included in analysis. Samples with a sequencing depth below 5000 counts were excluded from analyses. Only families seen with at least 3 counts in at least 10\% of samples were retained for subsequent analyses. 

\subsection{Priors}
The regression model required that 4 hyper-parameters $\Gamma$, $\Theta$, $\Xi$, and $\upsilon$  be specified.  We set $\Theta$ to a $D\times Q$ matrix of zeros representing our prior assumption that, on average, there was no association between each covarariate and microbial composition. We specified $\Gamma=I_Q$ to constrain associations between microbial composition and covariates to remain small. We specified  $\upsilon=D+3$ and $\Xi_{ij} = (\upsilon-D)$ if $i==j$ and $\Xi_{ij} = (\upsilon-D)/2$ if $i \neq j$ to reflect our weak prior assumption that the log absolute abundance of each taxa is uncorrelated \cite[p. 208-214]{aitchison1986}. 

\newpage
\beginsupplement
\section*{Supplementary Figures}

\begin{figure}[h]
    \centering
    \includegraphics{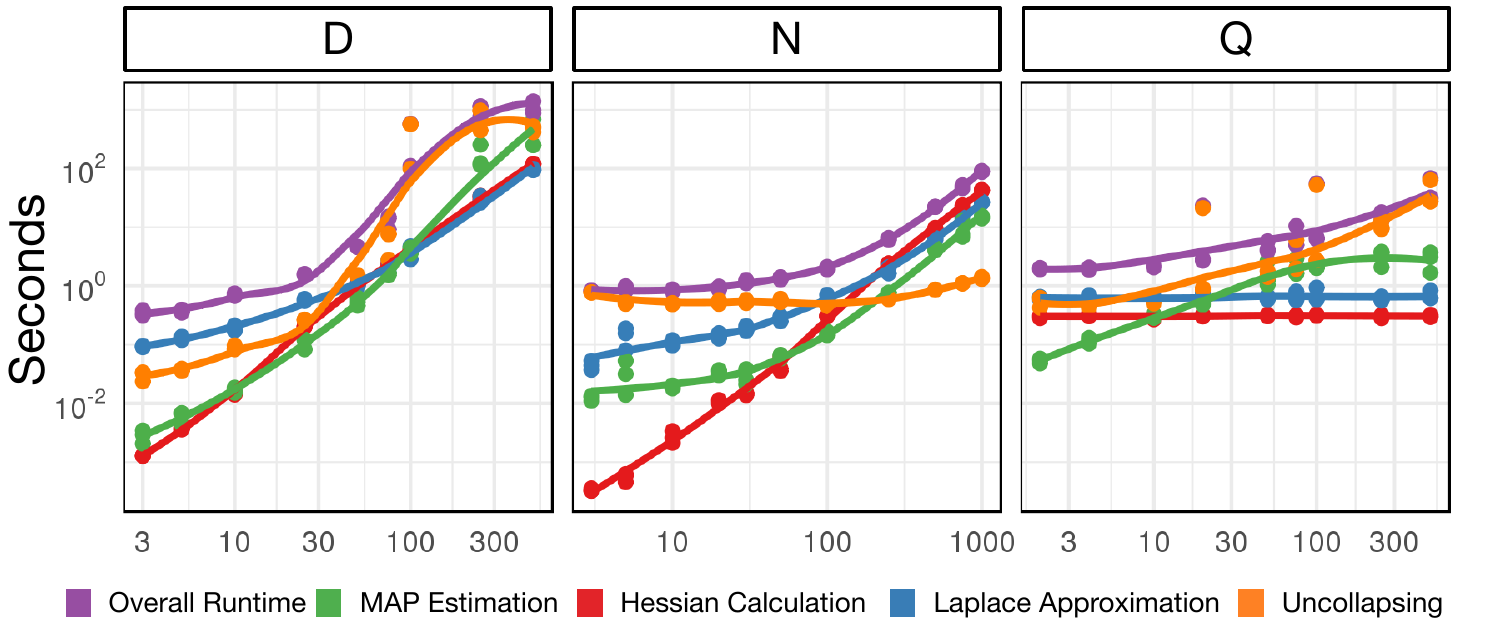}
    \caption[LA Collapsed (\textit{stray}) run-times decomposed into component processes.]{\textbf{LA Collapsed (\textit{stray}) run-times decomposed into component processes.}}
    \label{fig:runtimes}
\end{figure}

\begin{figure}[h]
    \centering
    \includegraphics[width=4in]{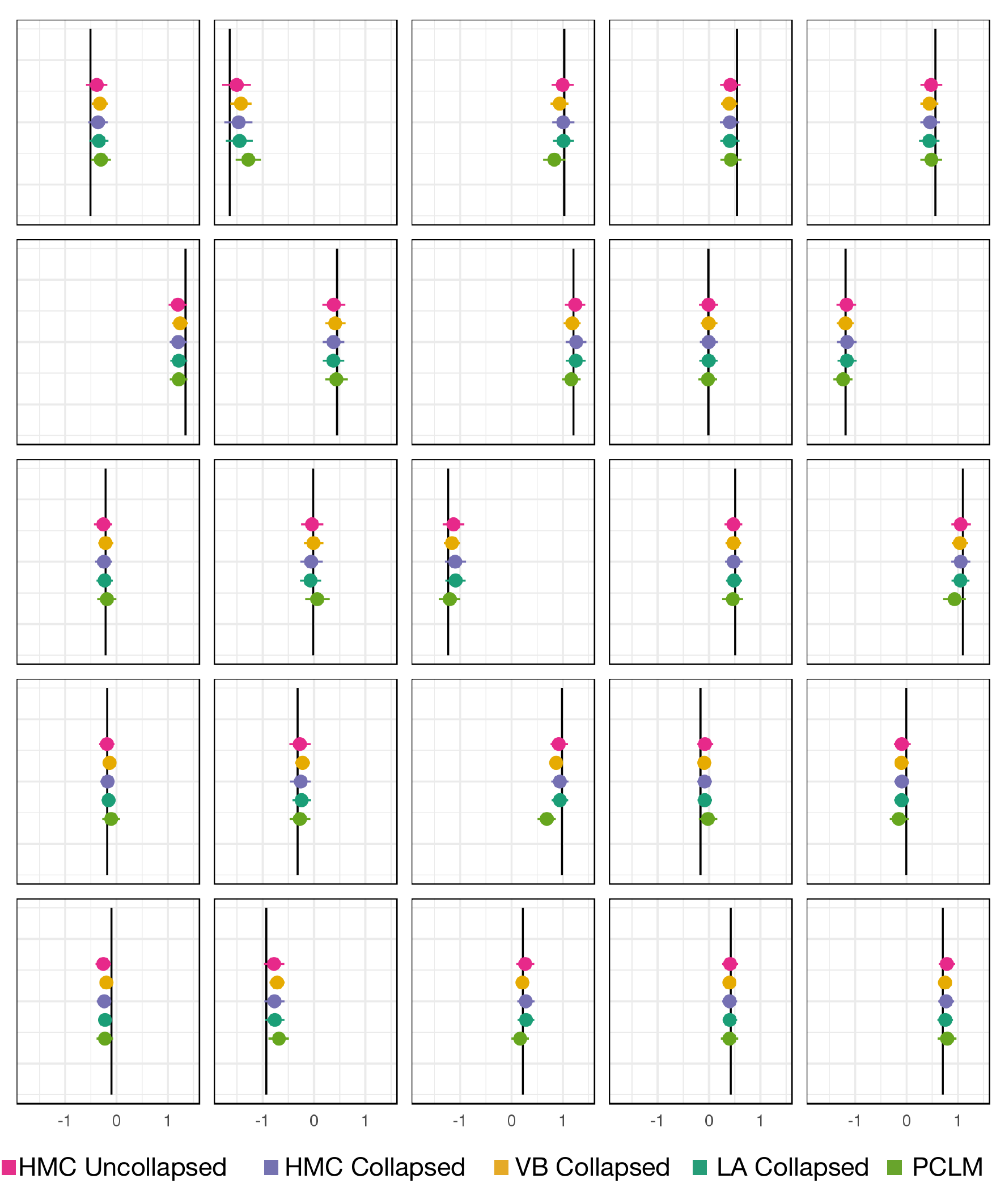}
    \caption[Example of simulation results in which uncertainty quantification from \textit{stray} is highly accurate]{\textbf{Example of simulation in which uncertainty quantification from LA Collapsed agrees with estimates from HMC.} For this simulation $N=30$, $D=30$, $Q=5$. Each panel represents a different element $\Lambda_{ij}$. The true simulated value of $\Lambda_{ij}$ in each panel is denoted by a black vertical line.}
    \label{fig:successcase}
\end{figure}

\begin{figure}[h]
    \centering
    \includegraphics[width=4in]{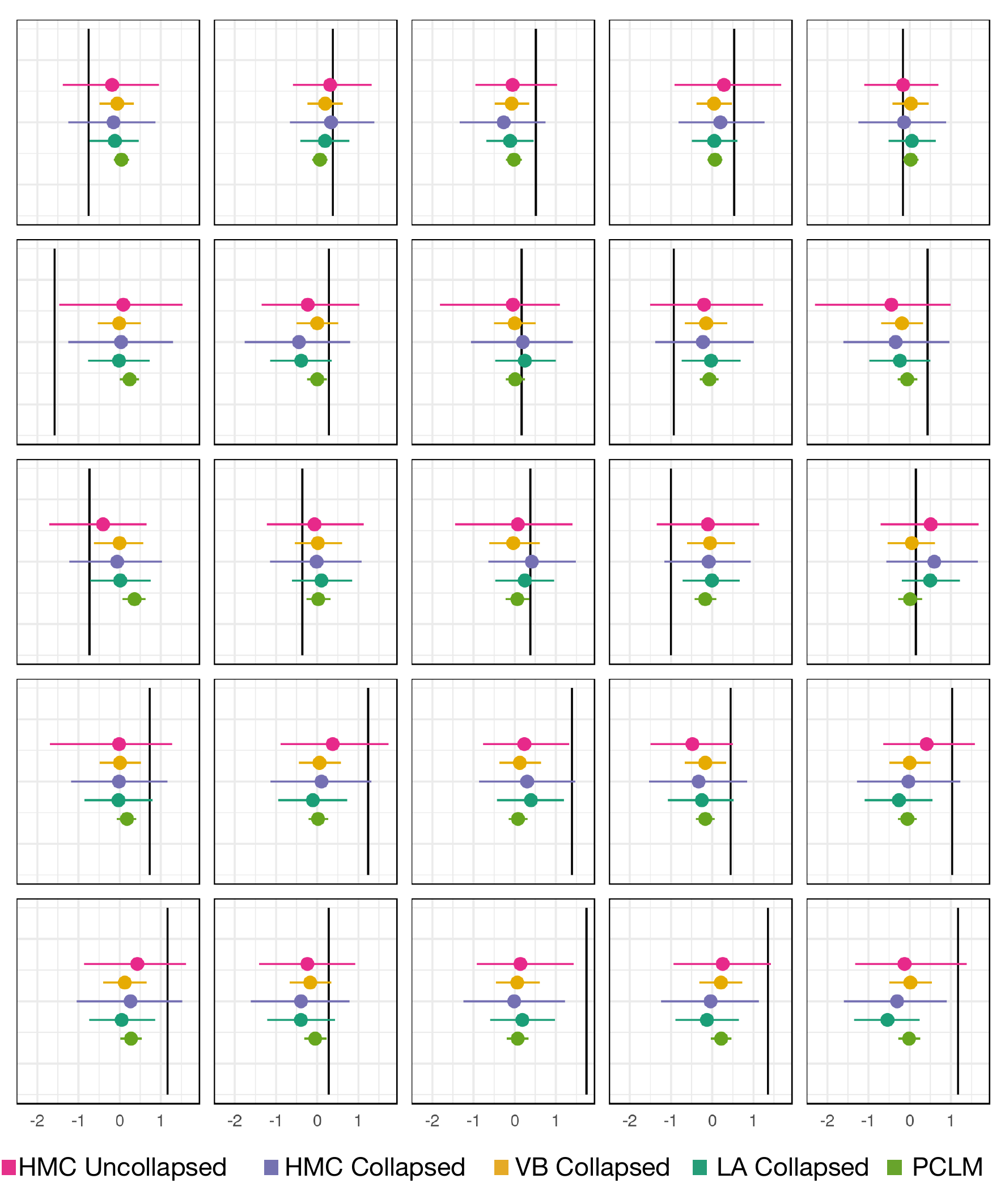}
    \caption[Example of simulation results in which uncertainty quantification from \textit{stray} disagrees with HMC estimates]{\textbf{Example of simulation in which uncertainty quantification from LA Collapsed disagrees with estimates from HMC.} For this simulation $N=100$, $D=30$, $Q=250$. Each panel represents a different element $\Lambda_{ij}$. The true simulated value of $\Lambda_{ij}$ in each panel is denoted by a black vertical line.}
    \label{fig:failcase}
\end{figure}

\begin{figure}[h]
    \centering
    \includegraphics{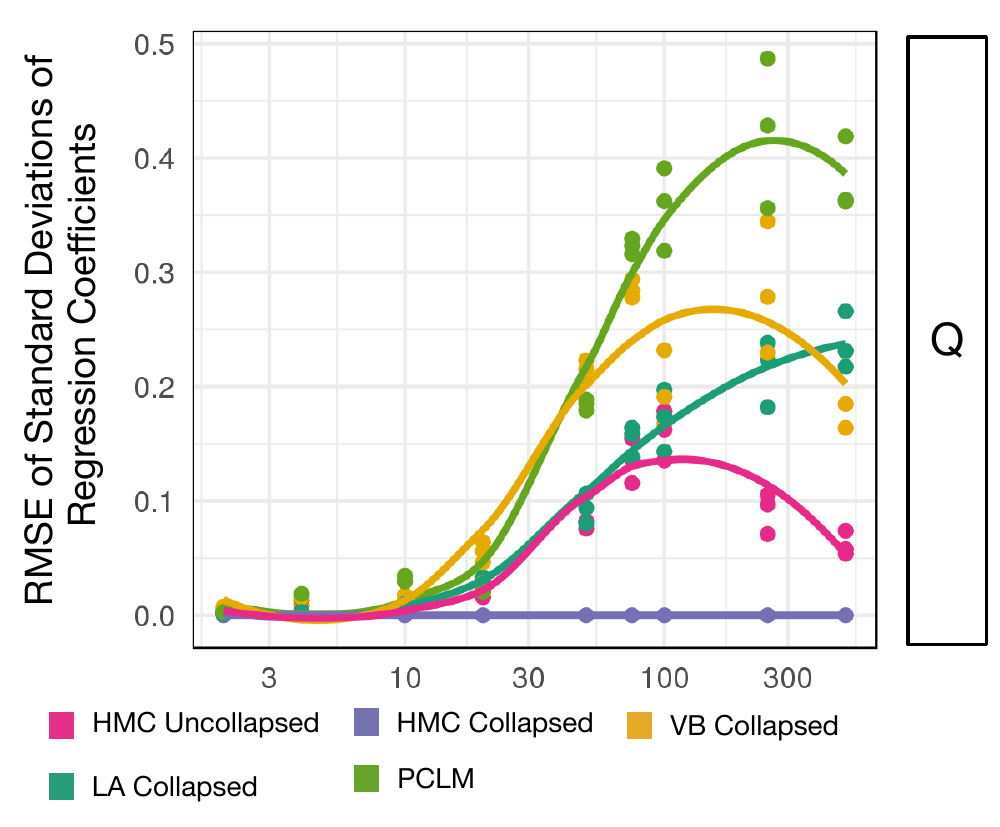}
    \caption[Contextualizing uncertainty quantification error using the Pseudo-Count Linear Model (PCLM)]{\textbf{Uncertainty quantification error can be contextualized by comparing against a fifth implementation (PCLM)}. The PCLM model consists of a pseudo-count based estimate of $\eta$ followed by the direct application of a multivariate conjugate linear model to estimate parameters $\Lambda$ and $\Sigma$. In this way the PCLM model ignores multinomial variation.}
    \label{fig:pclm}
\end{figure}

\begin{figure}[h]
    \centering
    \includegraphics{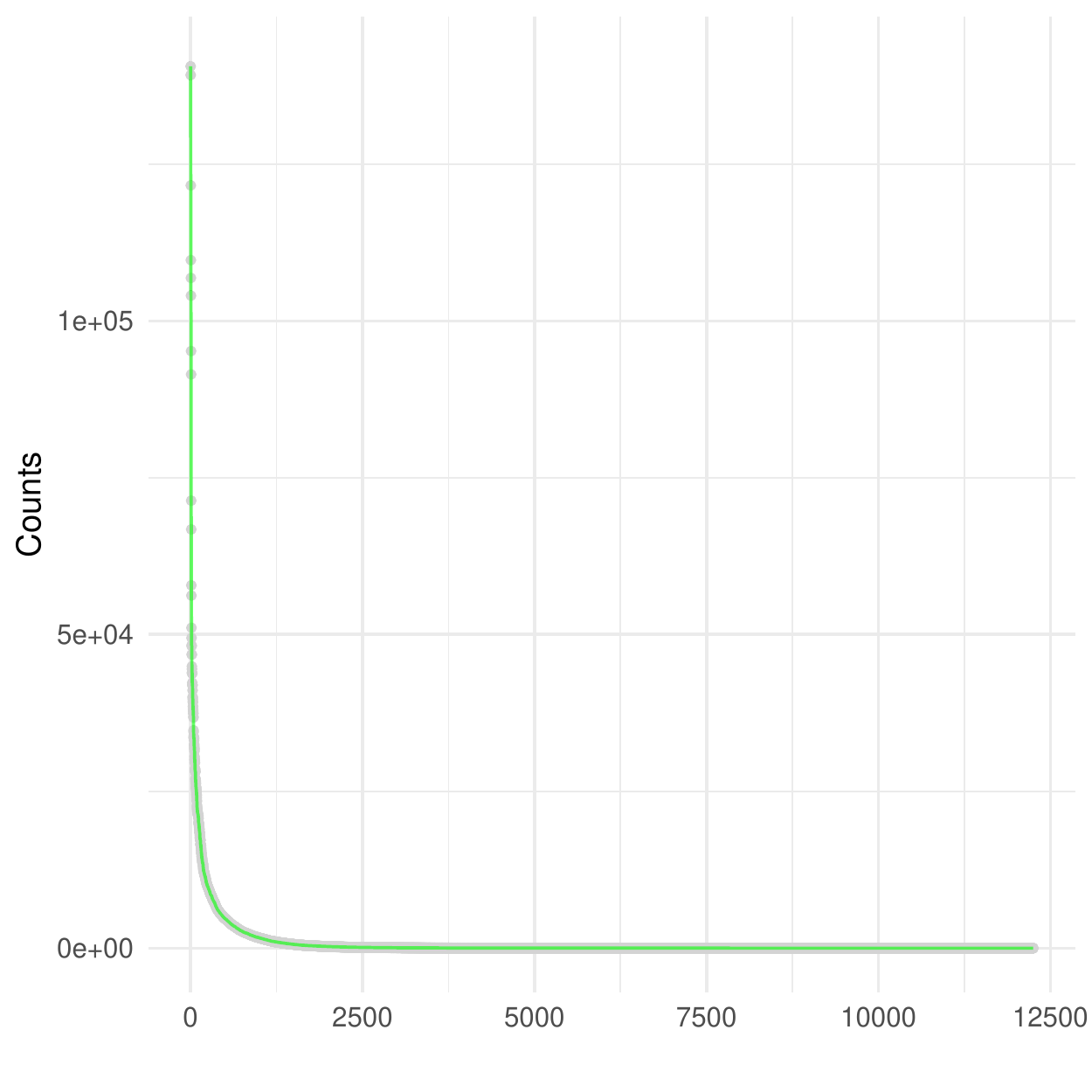}
    \caption[Posterior predictive checks of \textit{stray} applied to real data]{\textbf{Posterior predictive checks of \textit{stray} (\textit{i.e.,} LA Collapsed) applied to real data.} Each element of data $Y$ is ordered by value and denoted by a green line. The marginal posterior predictive distribution of each element is displayed based on its mean and 95\% credible interval in grey.}
    \label{fig:ppc}
\end{figure}

\begin{figure}[h]
    \centering
    \includegraphics{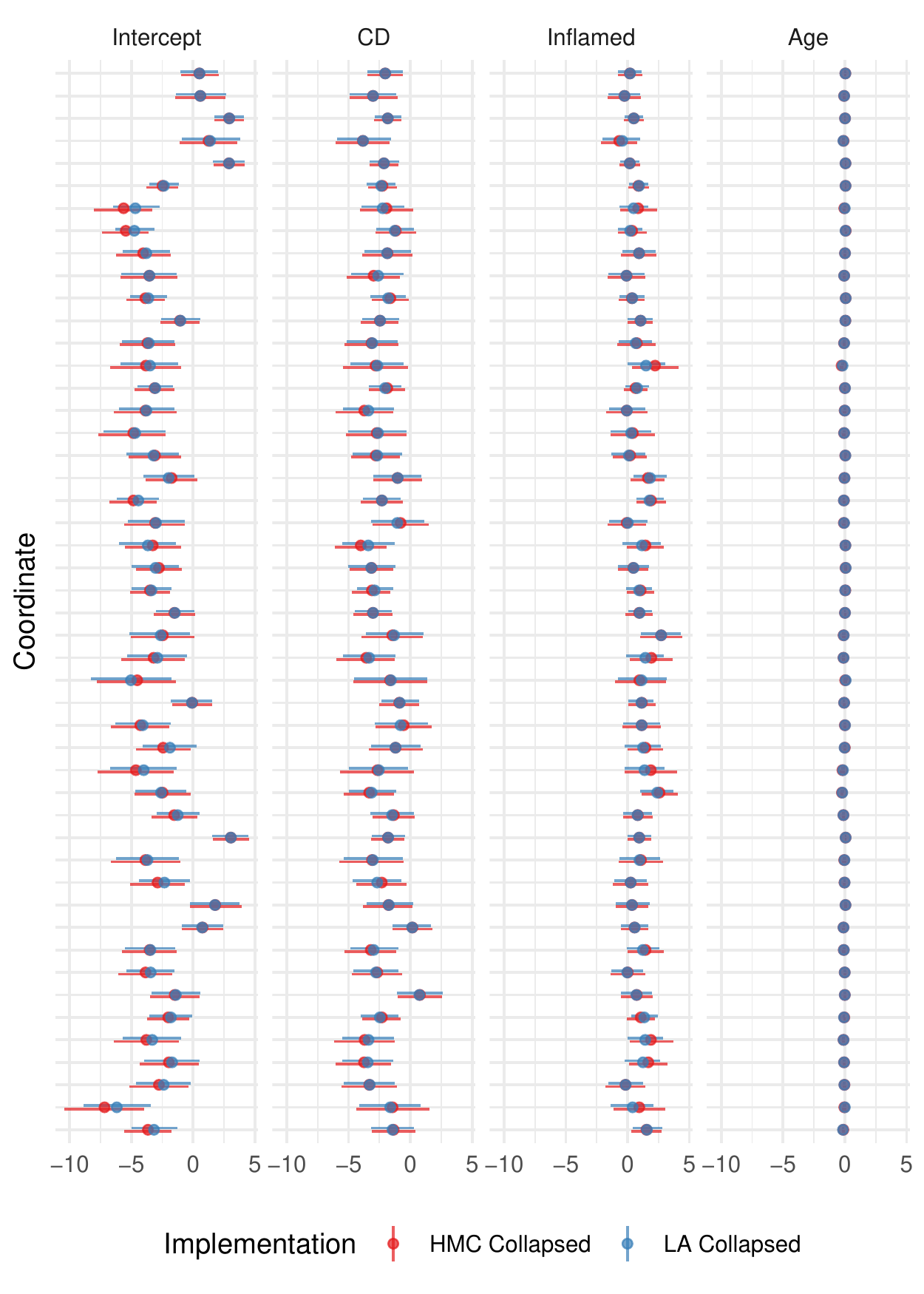}
    \caption[Comparison of posterior estimates of $\Lambda$ produced by different multinomial logistic-normal linear model implementations]{\textbf{Posterior estimates of $\Lambda$ produced by HMC Collapsed and LA Collapsed (\textit{stray}) are similar.} For each implementation the Posterior mean and 95\% credible interval is indicated.}
    \label{fig:real_data_comparision}
\end{figure}

\begin{figure}[h]
    \centering
    \includegraphics{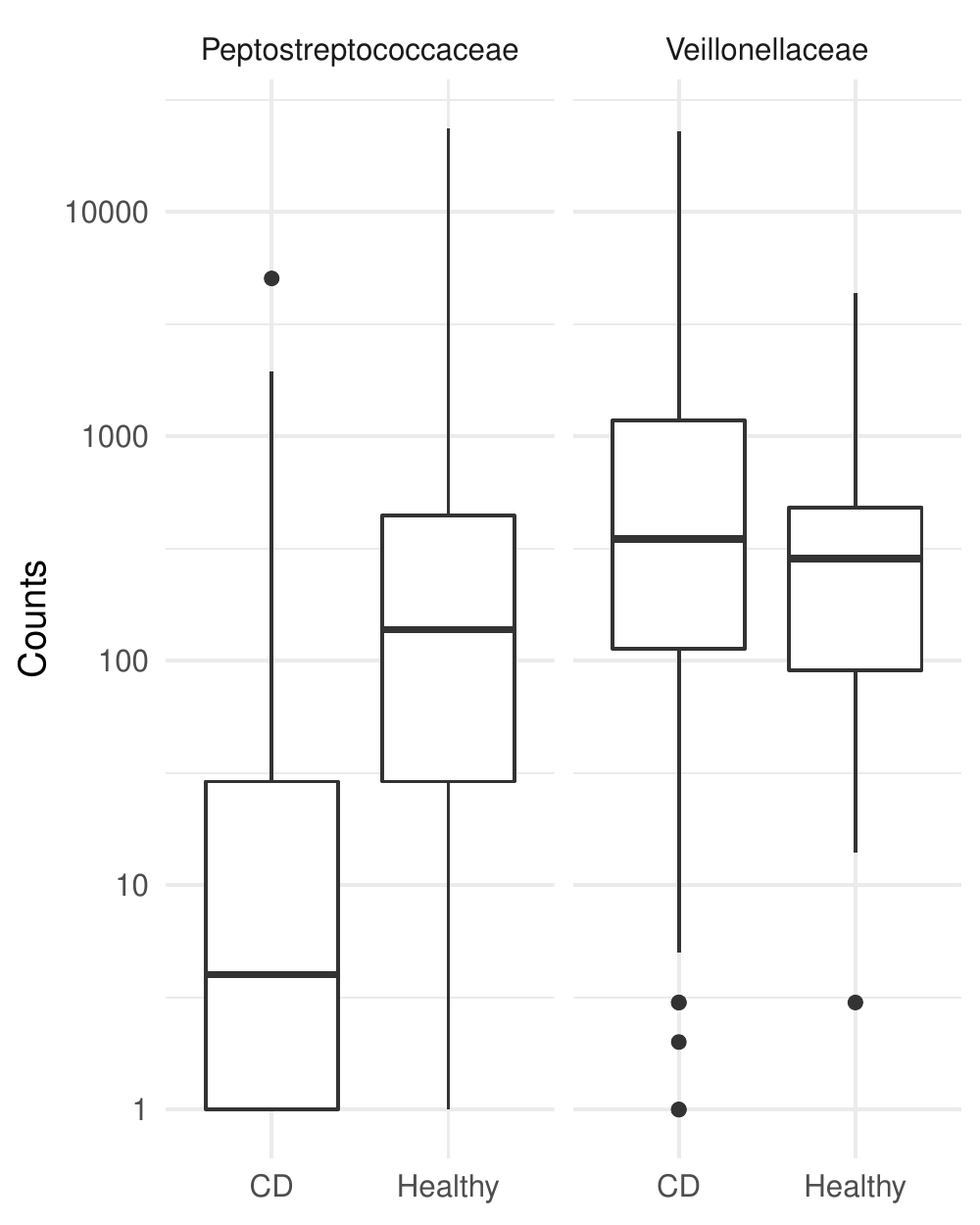}
    \caption[]{\textbf{Comparison of raw counts for CD and Healthy groups for Peptostreptococcaceae and Veillonellaceae families from Real Data analysis.} To allow visualization, a pseudo-count of 1 was added prior to log-transformation to avoid taking the log of zero. Boxplots show median, IQR, and $1.5\times$IQR.}
    \label{fig:compare_to_prior_analysis}
\end{figure}

\end{document}